
\tolerance=3000
\magnification=1100                      
\documentstyle{amsppt}  
\catcode`\@=11
\def\logo@{\relax}
\catcode`\@=\active
\loadbold

\def\Ran{\operatorname{Ran}}
\def\Ker{\operatorname{Ker}}

\def\im{\operatorname{im}}

\def\det{\operatorname{det}}

\def\vph{\varphi}

\def\a{\alpha}
\def\g{\gamma}

\def\b{\beta}
\def\s{\sigma}
\def\l{\lambda}

\document

\topmatter
\centerline
{\bf  ASYMPTOTIC STABILITY OF  MULTI-SOLITON SOLUTIONS}
\centerline
 {\bf FOR  NONLINEAR SCHR\"ODINGER EQUATIONS}

\head
Galina Perelman
\endhead

\centerline{Centre de Math\'ematiques}
\centerline{Ecole Polytechnique}
\centerline{F-91128 Palaiseau Cedex}
\centerline{France}
\centerline{e-mail: perelman\@math.polytechnique.fr}

\endtopmatter

\subsubhead Abstract
\endsubsubhead
We consider the Cauchy problem for the nonlinear Schr\"odinger equation
$i\psi_t=-\triangle\psi+F(|\psi|^2)\psi$,
  in space dimensions
$d\geq 3$, with initial data close to a sum of $N$
decoupled solitons. Under some suitable assumptions
on the spectral structure of
the one soliton linearizations we prove that for large time
the asymptotics of the solution is given by a sum of solitons with
slightly modified parameters and a  
small dispersive term.

\rm
\head
0. Introduction
\endhead 
In this paper we consider 
the nonlinear Schr\"odinger equation
$$
i\psi_t=-\triangle\psi+F(|\psi|^2)\psi, \quad (t,x)\in {\Bbb R}\times
{\Bbb R}^d,\,\, d\geq 3.\tag0.1$$
For suitable $F$  it possesses important  solutions of special form - 
solitary waves  (or, shortly, solitons):
$$ e^{i\Phi}\varphi(x-b(t),E),$$
$$\Phi=
\omega t+\gamma+{1\over 2}x\cdot v,\,\, b(t)=vt+c,\,\, 
E=\omega+{|v|^2\over 4}>0,$$
where $\omega,\,\gamma\in {\Bbb R}$,
$v,\,c\in{\Bbb R}^d$ are constants
and $\varphi$ is a ground state that is a smooth positive spherically
symmetric, exponentially decreasing solution of the equation
$$-\triangle\varphi+E\varphi+ F(\varphi^2)
\varphi=0.\tag 0.2$$
\par Solitary wave solutions are of special importance
not only because they are simple and sometimes explicit
solutions
of evolution equations, but also because of the distinguished
role they appear to play in the solution of the initial
value problem. This is best known for completely
integrable equations like
the cubic Schr\"odinger equation
$$
i\psi_t = -\psi_{xx} -|\psi|^2\psi.
\tag0.3$$ 
In the general position case the solution of the Cauchy problem
for this equation with rapidly decreasing smooth
initial data has in $  L_2({\Bbb R}) $
the asymptotic behavior 
$$ \psi \sim \sum_{j=1}^N e^{i\Phi_j}\vph(x-b_j(t),E_j)+ 
e^{il_0t}f_+, 
\quad \Phi_j=\omega_j t+\gamma_j+{xv_j\over 2},\,\,b_j=v_jt+c_j$$
where $ l_0= -\partial_x^2,$ and $f_+$ is some function in
 $ L_2({\Bbb R})$.  The number $ N,$
the function $ f_+ $ and the soliton parameters $ (\gamma_j,E_j,v_j,c_j)$ 
depend on the initial data. Due to possibility of 
explicitly integrating
equation (0.3) with the help of inverse scattering methods, 
they
can be described by effective formulas in terms of initial
condition. See, for example, [22] for these  results.
\par Numerical experiments have shown that even without
the presence
of an inverse-scattering theory, solutions, in general,
eventually resolve themselves into an approximate
superposition of weakly interacting solitary waves 
and decaying dispersive waves (see [11], for example). 
While exact theory confirming
 the special role
of solitary waves as a nonlinear basis with respect 
to which it is natural to view the solutions in the limit
of large time is not generally available,
partial indication is provided by stability theory
of such waves. A considerable literature 
has been devoted to the problem of orbital stability of
 solitons  following the work of Benjamin [1], 
see  also [7, 13, 14, 29, 34, 35]. The problem arises in connection with 
the Cauchy problem for equation (0.1) with initial data of the form
$$
\psi \big |_{t=0}=\vph(x,E_0) + \chi_0,
\tag0.4$$
where $\chi_0 $ is small in the Sobolev space
$ H^1({\Bbb R}^d) $. It was shown that under certain additional
conditions the solution $ \psi(x,t),\, t\geq 0 $ 
remains close (again in the space $ H^1({\Bbb R}^d) $)
to the surface 
$$\{e^{i\gamma}\vph(x-c,E_0),\,\gamma\in{\Bbb R},c\in{\Bbb R}^d\}.$$
 This notion of stability 
establishes that the shape of the wave is stable, 
but does not fully
resolve the question of what the asymptotic
 behavior of the system is.
\par  The 
first asymptotic stability results were
obtained by Soffer and  Weinstein
in the context of the equation
$$i\psi_t = 
-\triangle \psi+[V(x)+\lambda|\psi|^{m-1}]\psi, \tag0.5$$ 
(see [27, 28] and [30, 31, 32, 33, 36] for the further 
developments related to this model). 
The solitons for (0.5) arise as a perturbation
of the eigenfunction of the operator $-\triangle + V(x)$
and, in  contrast to the case of equation (0.1),
they  have a fixed center, which 
simplifies the analysis to some extent.

\par For the one- dimensional equation 
$$i\psi_t=-\psi_{xx}+F(|\psi|^2)\psi \tag0.6$$
the asymptotic stability of solitons was studied in
the works of Buslaev and author [4, 5]. We considered
the Cauchy problem (0.6), (0.4) and proved
that in the case where the spectrum of the linearization 
of equation (0.6) at the initial soliton has the simplest
 possible structure in some natural sense, 
the solution $ \psi $ has an asymptotic behavior of the form
$$
\psi=  e^{i\Phi_+}\vph(x-b_+(t),E_+)+ e^{-il_0t}f_++ o(1),
\quad \Phi_+=\omega_+ t+\gamma_++{xv_+\over 2},\,\,b_+=v_+t+c_+$$
as $ t\rightarrow +\infty $, where the parameters
$(\gamma_+,E_+,v_+c_+)$ of the limit soliton are close to the initial
ones $(0,E_0,0,0)$
and $ f_+ $ is small. Some  asymptotic results 
in the framework of significantly freer conditions on 
the linearization were obtained in [5], see also [6].
Recently the analysis of [4, 5, 6]  was extended to the multidimensional case
(0.1) by Cuccagna [8, 9]. 
\par 
As a natural generalization of the above situation one can consider
the case of several weakly interacting solitons. 
Assume that one has a set of solitons
$ e^{i\b_{0j}+i{x\cdot v_{0j}\over 2}}\vph(x-b_{0j},E_{0j})$, 
$j=1,\dots,N,$ 
that are well separated either in 
the original space or in Fourier space: for $j\neq k$,
either $|v^0_{jk}|$ or
$\min\limits_{t\geq 0}|b^0_{jk}(t)|$ is sufficiently large,
where $v^0_{jk}=v_{0j}-v_{0k}$, 
$b^0_{jk}(t)=b_{0j}-b_{0k}+v_{jk}^0t$.
In the second case we shall assume that the ``collision time'' $t^0_{jk}=-
{b_{jk}^0(0)\cdot v^0_{jk}\over |v^0_{jk}|^2}$
is ``bounded'' from above, see subsection 1.4, (1.7)
for the exact formulation.

 \par Consider the Cauchy problem
for  equation (0.1) with  initial data close to a sum
$$\sum\limits_{j=1}^N
e^{i\b_{0j}+i{x\cdot v_{0j}\over 2}}\vph(x-b_{0j},E_{0j}).$$
 If all the 
linearizations constructed independently
from the solitons $ \vph(E_{0j})$ satisfy
 the spectral conditions introduced in the case of one soliton, one 
can expect that as $ t \rightarrow +\infty $  
 the solution $ \psi $  looks like a sum of N soliton with slightly
modified parameters plus a small dispersive term.
In [23] 
this  was proved in the case $d=1,\, N=2$,
see also [19] for the asymptotic stability results  for the sums
of solitons in the context of KdV type equations.
The goal of the present paper is to extend the result of [23]
to the multidimensional case $d\geq 3$ (omitting also the
restriction N=2). 
The main new ingredient in the analysis is a combination
of the estimates for the
linear one soliton evolution obtained  by Cuccagna in
[8]  with the ideas of Hagedorn [15].
\par The structure of this paper is briefly as follows.
 It consists of two sections.
In the first section we introduce some preliminary objects and
state the main result.
The second contains  the complete proofs of the indicated results,
some technical details being removed to the appendices.

\head
1. Background and statement of the results
\endhead
\subhead
1.1. Assumptions on $F$
\endsubhead 
Consider the nonlinear Schr\"odinger equation
$$
i\psi_t=-\triangle\psi+F(|\psi|^2)\psi, \quad (t,x)\in {\Bbb R}\times
{\Bbb R}^d,\,\, d\geq 3.\tag1.1$$
We assume the following.
\proclaim{Hypothesis H0}
$F$ is a smooth function, $F(0)=0$,
 $F$ 
satisfies the estimates
$$ F(\xi)\geq -C\xi^q,\quad
|F^{(\alpha)}(\xi)|\leq C \xi^{p-\alpha},\,\,\, \a=0,1,2,$$
where $C>0$, $\xi\geq 1$, $q<{2\over d}$, $p<{2\over d-2}$.
\endproclaim 
Set
$g(\xi)=E\xi+F(\xi^2)\xi$.
\proclaim{Hypothesis H1}
\item{{\rm (i)}} There exists $\xi_0>0$ such that
$g(\xi)>0$ for $\xi<\xi_0$, $g(\xi)<0$ for $\xi>\xi_0$
and $g^\prime(\xi_0)<0$.
\item{{\rm (ii)}} There exists $\xi_1>0$ such that
$\int_0^{\xi_1} ds g(s)=0$.
\endproclaim
Further assumptions are given in terms of the
function
$$I(\xi,\lambda)=-\lambda \xi g^\prime(\xi)+(\lambda+2)g(\xi).$$
We consider $\xi_0$ of (H1) and assume: 
\proclaim{Hypothesis H2}
For any $\xi>\xi_0$ there exists a $\lambda(\xi)> 0$,
continuously depending on $\xi$, such that
$I(t,\lambda) \geq 0$ for $0<t<\xi$ and
$I(t,\lambda) \leq 0$ for $t>\xi$.
\endproclaim
We suppose hypotheses (H1,2)  to be true for $E$ in some open
interval 
${\Cal A}\subset{\Bbb R}_+$.
\par Under these assumptions equation (0.2) for $E\in {\Cal A}$,
has a unique 
positive spherically symmetric smooth exponentially decreasing solution
$\vph(x,E)$, see [2, 20]. More precisely, as $|x| \rightarrow \infty$
$$\vph(x,E)\sim Ce^{-\sqrt{E}|x|}|x|^{-{(d-1)\over 2}}.$$
This asymptotic estimate can be differentiated any number of 
times with respect to $x$ and $E$.
\par We shall call the functions
$w(x,\s)=\exp(i\beta+iv\cdot x/2)\vph(x-b,E)$,
$\s=(\beta,E,b,v)$ 
$\in{\Bbb R}^{2d+2}$
by soliton states. 
$w(x,\s(t))$ is a solitary wave solution iff $\s(t)$ satisfies the system:
$$\b^\prime=E-{|v|^2\over 4},\quad E^\prime=0,\quad
b^\prime=v,\quad v^\prime=0.\tag 1.2$$ 
\subhead
1.2. One soliton linearization
\endsubhead
Consider the linearization of equation (1.1) on a soliton $w(x,\s(t))$:
$$\psi \sim w+\chi,$$
$$i\chi_t=
(-\triangle +F(|w|^2))\chi+F^\prime(|w|^2)(|w|^2\chi+w^2\bar \chi).$$
Introducing the function $\vec f$:
$$ \vec f={f \choose \bar f},\quad \chi(x,t)= \exp(i\Phi)f(y,t),$$
$$\Phi=\beta(t) +{v \cdot x\over 2},\quad y=x-b(t),$$
one gets
$$i\vec f_t=L(E)\vec f,\quad L(E)=
L_0(E)+V(E),\,\, L_0(E)=(-\triangle+E)\s_3,$$
$$V(E)=V_1(E)\s_3+iV_2(E)\s_2,\quad
V_1=F(\vph^2)+F^\prime(\vph^2)\vph^2,\,\,V_2(E)=F^\prime(\vph^2)\vph^2.$$
Here $\s_2$, $\s_3$
are the standard Pauli matrices
$$\s_2=
\pmatrix0&-i\cr
i&0 \cr\endpmatrix,\quad
\s_3=
\pmatrix1&0\cr
0&-1 \cr\endpmatrix.$$
We consider $L$ as an operator in 
$L_2({\Bbb R}^d\rightarrow {\Bbb C}^2)$
defined on the domain where $L_0$ is self adjoint.
$L$ satisfies the relations
$$\s_3L\s_3=L^*,\quad \s_1L\s_1=-L,$$
where $\s_1=\pmatrix0&1\cr
1&0 \cr\endpmatrix $.
The continuous spectrum of $L(E)$ fills up two semi-axes
$(-\infty,E]$ and $[E,\infty)$. In addition 
$L(E)$ may have finite and finite dimensional point spectrum 
on the real and imaginary axis.
\par Zero is always  a point of the discrete spectrum. One can indicate 
$d+1$ eigenfunctions
$$\vec \xi_0=\vph{1\choose -1},\quad 
\vec \xi_j=\vph_{y_j}{1\choose 1},\,\,\, j=1,\dots d,$$
and $d+1$ generalized eigenfunctions
$$\vec \xi_{d+1}=-\vph_E{1\choose 1},\quad 
\vec \xi_{d+1+j}=-{1\over 2}y_j\vph{1\choose 1},\,\,\, j=1,\dots d,$$
$$L\vec\xi_j=0,\quad L\vec\xi_{d+1+j}=\vec\xi_j,\,\,\, j=0,\dots,d.$$
Let $M$ be the generalized null space of the operator $L$.
Under assumptions (H0,1,2), 
the vectors $\vec\xi_j,\,\,  j=0,\dots,2d+1$, span the subspace M iff 
$${d\over dE}\|\vph(E)\|_2^2\neq 0,$$
see [34, 20, 8].
\par We shall assume that
\proclaim{Hypothesis H3}
The set ${\Cal A}_0$ of $E\in {\Cal A}$ such that
\item{{\rm (i)}} zero is the only eigenvalue of the operator $L(E)$,
 and the dimension of the corresponding  generalized null space 
 is equal to $2d+2$;
\item {{\rm (ii)}} $\pm E$ {\it is not a resonance for} $L(E)$;
\newline is nonempty.
\endproclaim
Obviously, the set ${\Cal A}_0$ is open.
\par {\it Remark}.
$\pm E$ is said
to be a resonance of $L(E)$ if there is a solution
$\psi$ of the equation $(L(E)\mp E)\psi=0$ such that
$<x>^{-s}\psi \in L_2$ for any $s>1/2$ but not for $s=0$.
$\pm E$ can never be a resonance if 
$d\geq 5$, see lemma A4.3.

\par Consider the evolution operator $e^{-itL}$. One has the 
following proposition.
\proclaim{Proposition 1.1}  
For $E\in {\Cal A}_0$ and any $x_0,x_1 \in {\Bbb R}^d$,
$$\|\left<x-x_0\right >^{-\nu_0 }e^{-iL(E)t}\hat P(E)f\|_2
\leq C \left<t\right>^{-d/2}\|\left<x-x_1\right >^{\nu_0 }f\|_2,
\quad \nu_0>{d\over 2},\tag 1.3$$
where $\hat P(E)$ is the spectral projection  onto
the subspace of the continuous spectrum of $L(E)$:
$$\Ker \hat P=M,\quad \Ran \hat P=(\s_3M)^\bot.$$
The constant $C$ here is uniform
with respect to $x_0,x_1 \in {\Bbb R}^d$ and $E$ 
in  compact subsets of 
${\Cal A}_0$.
\endproclaim

This proposition 
is an immediate consequence of the $L_p$- $L_q$ 
estimates of
$e^{-iLt}\hat P$
proved by Cuccagna [8]. 
For the sake of completeness we sketch the proof
of (1.3) in appendix 4.

\subhead
1.3.
The nonlinear equation
\endsubhead We formulate here  the necessary facts about
the Cauchy problem for  equation (1.1)
with initial data in $H^1({\Bbb R}^d)$.
\proclaim{Proposition 1.2} 
Suppose that $F$ satisfies {\rm (H0)}. Then the Cauchy problem
for equation {\rm (1.1)} with initial data $\psi(x,0)=\psi_0(x)$,
$\psi_0\in H^1({\Bbb R}^d)$ has a unique solution $\psi$
in the space $C({\Bbb R}\rightarrow  H^1)$, and $\psi$ satisfies the
conservation laws
$$\int dx|\psi|^2=const,\quad H(\psi)\equiv
\int dx [|\nabla\psi|^2+U(|\psi|^2)]=const,$$
where $U(\xi)=\int_0^\xi dsF(s)$. Furthermore, for all $t\in {\Bbb R}$
$$\|\psi(t)\|_{H^1}\leq c(\|\psi_0\|_{H^1})\|\psi_0\|_{H^1},$$
where $c\,:\,{\Bbb R}_+\rightarrow {\Bbb R}_+$ is a smooth function.
\endproclaim
The assertion stated here can be found in [10, 11], for example.

\subhead
1.4. Description of the problem
\endsubhead
Consider the Cauchy problem for equation (1.1) with initial data
$$\psi|_{t=0}=\psi_0\in H^{1}\cap L_1 ,\quad 
\psi_0=\sum\limits_{j=1}^Nw(\cdot,\s_{0j})+\chi_0,\tag 1.4$$
$$\s_{0j}=(\b_{0j},E_{0j},b_{0j},v_{0j}),
\quad \min_{j\neq k}|v^0_{jk}|\geq v_0>0.\tag1.5$$
Here $v^0_{jk}=v_{0j}-v_{0k}$.
Set $b_{jk}^0=b_{0j}-b_{0k}$, $j\neq k$. Write
$b_{jk}^0$ as the sum
$$b_{jk}^0=r_{jk}^0-t_{jk}^0v^0_{jk},
\quad r_{jk}^0\cdot v_{jk}^0=0,\,\,
t_{jk}^0=-{b_{jk}^0\cdot v^0_{jk}\over |v^0_{jk}|^2}. \tag 1.6$$
For $j\neq k$ we define the effective small parameter
$\epsilon_{jk}$:
$$\epsilon_{jk}=\cases
(\min\limits_{t\geq 0}|b^0_{jk}(t)|+|v^0_{jk}|)^{-1},\,\,
\roman{if} \,\, t_{jk}^0\leq \kappa< r_{jk}^0>,\cr
|v^0_{jk}|^{-1}\quad \roman{otherwise},\cr\endcases\tag 1.7$$
where $b^0_{jk}(t)=b_{jk}^0+tv^0_{jk}$,
$\kappa$ is a fixed positive constant.
\par Assume that
\item {(T1)}  $\epsilon\equiv\max\limits_{j\neq k}\epsilon_{jk}$ 
{\it is sufficiently small}\footnote{``Sufficiently small (large)''
assumes  constants that depend only on $v_0$,
$\kappa$ and $E_{0j},j=1,\dots,N $.};
\item {(T2)} $E_{0j}\in {\Cal A}_0$, $j=1,\dots, N$.
\newline
Our goal is to describe the asymptotic behavior of the solution $\psi$
as $t\rightarrow +\infty$, provided $\chi_0$ is sufficiently small
in the following sense:
\item {(T3)}  {\it for some $m^\prime$, 
${1\over m}+{1\over m^\prime}=1$, $m\geq 2p+2$,
${4\over d}+2<m<{4\over d-2}+2$ if $d\geq 4$ ,
$4\leq m<{4\over d-2}+2$ if $d=3$,
 the norm
$${\Cal N}=\|\chi_0\|_1+\|\hat \chi_0\|_{m^\prime}$$
is sufficiently small}.
\newline
Here $\hat \chi_0$ stands for the Fourier transform of $\chi_0$.
\par Our main result is given by the following theorem.
\proclaim{Theorem 1.1}
For $t\geq 0$  the solution $\psi$ of {\rm (1.1), (1.4)}
admits the representation
$$ \psi(t)=\sum\limits_{j=1}^Nw(\cdot,\s_{j}(t))+\chi(t),\quad
\s_j(t)=(\b_j(t),E_j(t),b_j(t),v_j(t)),$$
where $|E_j(t)-E_{0j}|$, $|v_j(t)-v_{0j}|$, $j=1,\dots, N$,
$\|\chi(t)\|_{L_2\cap L_m}$ are small  uniformly w.r.t.
$t\geq 0$, and as $t\rightarrow +\infty$,
$$ \|\chi(t)\|_{m}=O(t^{-d({1\over 2}-{1\over m})}).$$
Moreover, there
exist vectors $\s_{+j}=(\b_{+j}, E_{+j}, b_{+j}, v_{+j})$,
such that as $t\rightarrow +\infty$,
$$|\s_j(t)-\s_{+j}(t)|=O(t^{-\delta}),$$
for some $\delta>0$. Here $\s_{+j}(t)$
is the trajectory of {\rm (1.2)} with the initial data
$\s_{+j}(0)=\s_{+j}$.
\endproclaim

\head
2. Proof of the theorem 
\endhead

Up to some technical modifications 
the main line of the proof repeats that of [23].
\subhead
2.1. Splitting of the motions
\endsubhead
Following [23] we decompose the solution $\psi$ as follows.
$$\psi(x,t)=\sum\limits_{j=1}^Nw(x,\s_{j}(t))+\chi(x,t).\tag2.1$$
Here $\s_j(t)=(\b_j(t),E_j(t),b_j(t),v_j(t))$ is an arbitrary 
trajectory in the set of admissible values of parameters, 
it is not a solution of (1.2) in general.
\par We fix the decomposition (2.1) by imposing the orthogonality conditions
$$\left<\vec f_j(t),\s_3\vec \xi_k(E_j(t))\right>=0,
\quad j=1,\dots,N,\,\,\,k=0,\dots,2d+1.\tag2.2$$
Here
$$\vec f_j={f_j\choose \bar f_j},\quad 
\chi(x,t)=\exp(i\Phi_j)f_j(y_j,t),$$
$$\Phi_j=\b_j(t)+v_j\cdot x/2,\quad y_j=x-b_j(t),$$
$<\cdot,\cdot>$ is the  inner product in
$L_2({\Bbb R}^d\rightarrow {\Bbb C}^2)$.
\par Geometrically these conditions mean that for each t the vector 
$\vec f_j(t)$ belongs to the subspace of the continuous spectrum of 
the operator $L(E_j(t))$.
\par For $\psi$ of the form (1.4) with
$\min\limits_{{j,k\atop j\neq k}}(|v^0_{jk}|+|b^0_{jk}|)$ sufficiently large,
 and with
$\chi_0$ sufficiently small in
some $L_p$ norm, 
the solvability of (2.2) is guaranteed 
by the non-degeneration of the corresponding Jacobi matrix,
see lemma A1.1.
So, one can assume that the initial decomposition (1.4)
obeys (2.2). To prove the existence of a decomposition 
(2.1), (2.2) for
$t>0$, 
one can invoke a standard continuity type argument,
 see appendix 1 for the details.
\par Rewriting (2.1) as an equation for $\chi$ one gets
$$ i\vec\chi_t=H(\vec\s(t))\vec\chi+N,
\tag 2.3$$
Here 
$$ \vec \chi={\chi\choose\bar \chi},\quad 
\vec \s=(\s_1,\dots,\s_N)\in {\Bbb R}^{(2d+2)N},$$
$$ H(\vec\s)=-\triangle\s_3+\sum\limits_{j=1}^N {\Cal V}(w_j),$$
$${\Cal V}(w)=(F(|w|^2)+F^\prime(|w|^2)|w|^2)\s_3+
F^\prime(|w|^2)\pmatrix 0&w^2\cr
-\bar w^2&0\cr\endpmatrix,\quad w_j=w(x,\s_j).$$
The nonlinearity $N$ is given by the following expression
$$N=N_0+\sum\limits_{j=1}^Ne^{i\s_3\Phi_j}l(\s_j)\vec\xi_0(y_j,E_j),$$
$$ N_0=F(|\psi_s+\chi|^2){\psi_s+\chi\choose
-\bar \psi_s-\bar \chi}-$$
$$\sum\limits_{j=1}^N\left(F(|w_j|^2){w_j\choose-\bar w_j}+
{\Cal V}(w_j)\vec \chi\right),\quad 
\psi_s=\sum\limits_{j=1}^N w_j,$$
$$l(\s_j)=\g^\prime_j
+{1\over 2}v_j^\prime\cdot y_j+
ic_j^\prime\cdot\nabla\s_3-iE_j^\prime\partial_E\s_3,$$
where $\g_j$, $c_j$ are defined as follows.
$$\b_j(t)=\int_0^t ds (E_j(s)-
{|v_j(s)|^2\over 4}-{v_j^\prime(s)\cdot b_j(s)\over 2})+\g_j(t),
\quad b_j(t)=\int_0^t ds v_j(s)+c_j(t).$$
In terms of parameters $(\g,E,c,v)$ (1.2) takes the form
$$\g^\prime=0,\,\, E^\prime=0,\,\,c^\prime=0,\,\,v^\prime=0.$$ 
\par Substituting the expression for $\chi_t$ from 
(2.3) into the derivative of the orthogonality conditions, one gets
for $j=1,\dots,N$
$$ie(E_j)E_j^\prime=
\left<N_j,\s_3e^{i\Phi_j}\vec\xi_0(\cdot-b_j,E_j)\right>+
\left<\vec f_j,l(\s_j)\vec\xi_0(E_j)\right>,$$
$$n(E_j)v_j^\prime=
\left(\left<N_j,\s_3e^{i\Phi_j}\vec\xi_k(\cdot-b_j,E_j)\right>+
\left<\vec f_j,l(\s_j)\vec\xi_k(E_j)\right>\right)_{k=1,\dots,d},$$
$$e(E_j)\g_j^\prime=
\left<N_j,\s_3e^{i\Phi_j}\vec\xi_{d+1}(\cdot-b_j,E_j)\right>+
\left<\vec f_j,l(\s_j)\vec\xi_{d+1}(E_j)\right>,\tag 2.4$$
$$in(E_j)c_j^\prime=-
\left(\left<N_j,\s_3e^{i\Phi_j}\vec\xi_{d+1+k}(\cdot-b_j,E_j)\right>+
\left<\vec f_j,l(\s_j)\vec\xi_{d+1+k}(E_j)\right>\right)_{k=1,\dots,d}.$$
Here $$N_j=N_0+\sum\limits_{k, k\neq j}{\Cal V}(w_k)\vec \chi+
\sum\limits_{k,k\neq j}e^{i\s_3\Phi_k}l(\s_k)\vec\xi_0(y_k,E_k),
\quad j=1,\dots,N,$$
$$e={d\over dE}\|\vph\|_2^2,\,\,
n={1\over 2}\|\vph\|_2^2.$$
The right hand side of (2.4) 
also contain the derivative $\vec \s^\prime$, 
which enters linearly in $l(\s_k)$. In principle, system (2.4) 
can be solved with respect
to derivative and together with equation 
(2.3) constitutes a complete system for
$\vec \s$ and $\chi $:
$$ i\vec\chi_t=H(\vec\s(t))\vec\chi+N(\vec \s, \vec \chi),\tag 2.5$$
$$
\vec\s^\prime=G(\vec\s,\vec\chi),\quad
\chi|_{t=0}=\chi_0,\quad \s_j(0)=\s_{0j}.\tag 2.6$$
\subhead
2.2. Integral representations for $\chi$
\endsubhead
In this subsection we follow closely the constructions 
of  Hagedorn [15] (developed  in order to  prove
the asymptotic completeness for the charge transfer model),
see also [21]. We start by 
rewriting (2.5) as an integral equation
$$\vec\chi(t)={\Cal U}_0(t,0)\chi_0-i
\int_0^t{\Cal U}_0(t,s)\left[\sum\limits_{j=1}^N{\Cal V}_j(s)
\vec\chi(s)
+N\right]
ds,\tag 2.7$$
Here ${\Cal U}_0(t,\tau)=e^{i(t-\tau)\triangle \s_3}$,
${\Cal V}_j={\Cal V}(w_j)$.
\par Next we introduce the one soliton adiabatic propagators
${\Cal U}_j^A(t,\tau)$:
$$i{\Cal U}_{j\,t}^A(t,\tau)=L_j(t){\Cal U}_j^A(t,\tau),\quad 
{\Cal U}_j^A(t,\tau)|_{t=\tau}=I,$$
$$L_j(t)=-\triangle\s_3+\tilde{\Cal V}_j(t)+
R_j(t),\quad R_j(t)=iT_{0j}(t)[ P_j^\prime(t),P_j(t)]T_{0j}^*(t),$$
$$\tilde{\Cal V}_j(t)=
T_{0j}(t)T_{j}(t)V(E_{0j})T_{j}^*(t)T_{0j}^*(t),\quad 
P_j(t)=T_{j}(t)\hat P(E_{0j})T_{j}^*(t).$$
Here
$$T_{0j}(t)=B_{\b_{0j}(t), b_{0j}(t), v_{0j}},\quad 
T_{j}(t)=B_{\theta_{j}(t), a_{j}(t), 0},$$
$$\theta_j=\int_0^t ds \left (E_j(s)-E_{0j}+{|v_{j}(s)-v_{0j}|^2\over 4}
\right),\quad a_j=\int_0^t ds (v_j(s)-v_{0j}),$$
$$(B_{\b,b,v}f)(x)=e^{i\b\s_3+i{v\cdot x\over 2}\s_3}f(x-b),$$
$\s_{0j}(t)=(\beta_{0j}(t),E_{0j},b_{0j}(t), v_{0j})$ 
being the solution of (1.2) with initial data
$\s_{0j}(0)=\s_{0j}$.
Obviously,
$$ P_j^A(t){\Cal U}_j^A(t,\tau)={\Cal U}_j^A(t,\tau) P_j^A(\tau),$$
where
$$ P_j^A(t)=T_{0j}(t)P_j(t)T_{0j}^*(t).$$
Write the solution $\chi$ as the sum:
$$\vec\chi(t)=\vec h_j(t)+\vec k_j(t), \quad 
\vec h_j(t)= P_j^A(t)\vec\chi(t).$$

Using the adiabatic evolution ${\Cal U}_j^A(t,\tau)$
 one can write the following representation for $h_j(t)$
$$
\vec h_j(t)={\Cal U}_j^A(t,0)P_{j}^A(0)\vec\chi_0-
i\int_0^t{\Cal U}_{j}^A(t,s)P_{j}^A(s)
[\sum\limits_{m,\,m\neq j}{\Cal V}_{m}(s)\vec\chi(s)+D_j(s)]ds,\tag 2.8$$
Here 
$$D_j=N+({\Cal V}_j-\tilde{\Cal V}_j)\vec\chi-R_j\vec\chi.\tag 2.9$$

Combining (2.7), (2.8) one gets finally
$$\vec\chi=\roman{(I)+(II)+(III)+(IV)},\tag 2.10$$
where
$$\roman{(I)}={\Cal U}_0(t,0)\vec\chi_0-i\sum\limits_{j}
\int_0^t ds{\Cal U}_0(t,s)
\tilde{\Cal V}_{j}(s){\Cal U}_{j}^A(s,0)P_{j}^A(0)\vec\chi_0,$$
$$\roman{(II)}=-
\sum\limits_{{j,m\atop j\neq m}}\int_0^t dsK_j(t,s)
{\Cal V}_{m}(s)\vec\chi(s),$$
$$\roman{(III)}=-i\int_0^tds{\Cal U}_0(t,s)D,$$
$$\roman{(IV)}=-
\sum\limits_{j}\int_0^t dsK_j(t,s)D_j(s).$$
Here
$$ D=N+
\sum\limits_{j}\left(\tilde{\Cal V}_{j}\vec k_j
+({\Cal V}_{j}-\tilde{\Cal V}_{j})\vec\chi\right),\tag 2.11$$
$$K_j(t,s)=\int_s^t d\rho{\Cal U}_0(t,\rho)
\tilde{\Cal V}_{j}(\rho){\Cal U}_{j}^A(\rho,s)P_{j}^A(s).$$

The relations (2.4), (2.7), (2.10) 
make up the final form of the equation 
which is used to prove theorem 1.1. 
\subhead
2.3. Estimates of solitons parameters
\endsubhead
Following [4, 23] we consider (2.4), (2.7), (2.10) 
on some finite interval $[0,t_1]$ and then study 
the limit $t_1\rightarrow +\infty$. On the interval
$[0,t_1]$ we 
introduce a natural system of norms for the components of the
solution $\psi$:
$$M_0(t)=\sum\limits_{j=1}^N
|\g_j(t)-\b_{0j}|+
|E_j(t)-E_{j0}|+|c_j(t)-b_{0j}|+|v_j(t)-v_{0j}|,$$
$$M_1(t)=\sum_{j=1}^N\|<y_j>^{-\nu}\chi(t)\|_2,
\quad M_2(t)=\|\chi(t)\|_{2p+2},\quad \nu>{d+2\over 2},$$
without loss of generality one can assume that
$m=2p+2$.
\par These norms generate the system of majorants
$${\Bbb M}_0(t)=\sup\limits_{0\leq\tau\leq t}M_0(\tau),\,\,
{\Bbb M}_l(t)=\sup\limits_{0\leq\tau\leq t}M_l(\tau)\rho^{-\mu_l}(\tau),
\,\, l=1,2,\quad \hat {\Bbb M}_k={\Bbb M}_k(t_1).$$
Here 
$1<\mu_1<{3\over 2}$ if $d=3$ and $1<\mu_1={dp\over 2}$ for
$d\geq 4$,
$\mu_2=d({1\over 2}-{1\over 2p+2})$,
$$\rho(t)=<t>^{-1}+\sum\limits_{{j,k\atop j\neq k}}<t-t_{jk}>^{-1},$$
$t_{jk}$ being ``the collision times'' that are defined as follows.
We set $t_{jk}=0$ if $t_{jk}^0\leq 0$. For $(j,k)$ such that
 $t_{jk}^0>0$, we define
$t_{jk}$  by the relation,
$$\int_0^{t_{jk}}ds{\tilde v_{jk}(s)\cdot v_{jk}^0\over|v_{jk}^0|^2}
=t_{jk}^0,$$
where
$$\tilde v_{jk}(t)=
\cases v_{jk}(t),\,\, \roman{if}\,\, t\leq t_1,\cr
v_{jk}(t_1),\,\, \roman{if}\,\, t> t_1,\cr\endcases
\quad v_{jk}(t)=v_j(t)-v_k(t).$$
 Let us mention that
\par (i) $t_{jk}$ are well defined 
provided  
$| v_{jk}(t)-v_{jk}^0|< v_0$, $0\leq t\leq t_1$;
\par (ii) the collision times $t_{jk}$ belonging to the interval 
$[0,t_1]$ ``do not depend on $t_1$''.

\par It follows directly from the definition  of $M_0$ that
$$|\theta_j^\prime(t)|,\, |a_j^\prime(t)|\leq M_0(t)+
 M_0^2(t),\,\,\,
|b_j(t)-\tilde b_j(t)|\leq  M_0(t),\tag2.12$$
$$|\Phi_j(x,t)-\tilde\Phi_j(x,t)| \leq
 M_0(t)<x-b_j(t)> +{\Bbb M_0}(t)\int_0^tds |c_j^\prime(s)|,\tag2.13$$
where
$$\tilde b_j(t)=b_{0j}(t)+a_j(t),
\quad 
\tilde\Phi_j(x,t)=\beta_{0j}(t)+\theta_j(t)+v_{0j}\cdot x/2.$$
It is also easy to check that $\tilde b_{jk}=\tilde b_j-\tilde b_k$
 admits the estimates
$$|\tilde b_{jk}(t)|\geq c|v^0_{jk}||t-t_{jk}|,\tag 2.14$$
$$|\tilde b_{jk}(t)|\geq c(\min\limits_{s\geq 0}|b^0_{jk}(s)|
+|v^0_{jk}||t-t_{jk}|)-c ,\quad t_{jk}^0\leq \kappa <r_{jk}^0>\tag 2.15$$
provided $ M_0(t)\leq c$ for $0\leq t\leq t_1$.
 Here and below $c$ is used as a general notation of
positive constants that depend only on $v_0, \kappa$
and eventually on $ E_j,
j=1,\dots,N$, in that case they  can be chosen 
uniformly with respect to 
$E_j$ in some finite vicinity of $E_{0j}$.

\par Consider relations (2.4).
Since 
$$|N_0|\leq c\left(\sum\limits_{{j,k\atop j\neq k}}
|w_j||w_k|(1+|\chi|)+
\cases 
|\chi|^2 +|\chi|^{2p+1} \,\,\,\,\roman{if}\,\, 
p>{1\over 2},\cr
|\chi|^{2} \quad\quad\,\,\roman{if}\,\, 
p\leq {1\over 2}\cr\endcases\right)$$
and 
$$ |<e^{i\s_3\Phi_j}l(\s_j)\vec\xi_0(\cdot-b_j,E_j),
\s_3e^{i\s_3\Phi_k}\vec \xi_l(\cdot-b_k,E_k)>|=
O(|\lambda_j|e^{-c |b_{jk}|}|v_{jk}|^{-\infty}),$$
$ j\neq k$,
$ \l_j=
(\g_j^\prime,E_j^\prime,c_j^\prime,v_j^\prime)$,
$b_{jk}=b_j-b_k$,
one gets immediately from (2.4)
$$|\l_j(t)|\leq W({\Bbb M})
[\sum\limits_{{i,l\atop i\neq k}}e^{-c|b_{ik}(t)|}+\left({\Bbb M}_1^2(t)+
{\Bbb M}_2^{2}(t)
\right)\rho^{2\mu_1}(t)].\tag2.16$$
\par We use $W({\Bbb M})$ as a general notation for
  functions of ${\Bbb M}_0$, 
${\Bbb M}_1$, ${\Bbb M}_2$, which are bounded in some finite vicinity
of the point ${\Bbb M}_l=0,\,l=0,1,2,$ and may acquire $+\infty$ out 
some larger vicinity. They depend only on 
$v_0$, $\kappa_0$, $E_{j0},\,\,j=1,\dots,N$
and can be chosen to be spherically 
symmetric and monotone.
In all the formulas where $W$ appear
it would not be hard to replace them by  some explicit expressions
but such expressions are useless for our aims.

\par Combining (2.13), (2.16) one gets
$$|\Phi_j(x,t)-\tilde\Phi_j(x,t)| \leq W({\Bbb M})
 {\Bbb M_0}(t)<x-b_j(t)>.\tag 2.17$$
\par Integrating (2.16) and taking into account (2.14), (2.15) we  obtain
$${\Bbb M}_0\leq W(\hat{\Bbb M})[\epsilon+{\Bbb M}_1^2+
{\Bbb M}_2^{2}].\tag 2.18$$
\par Consider the vectors $\vec k_j(t)=(I-P_j^A(t))\vec \chi(t)$,
$\vec k_j(x,t)=\sum_{l=0}^{2d+1}k_{jl}(t)e^{i\tilde \Phi_j\s_3}
\vec \xi_{l}(x-\tilde b_j(t),E_{0j})$.
The orthogonality
conditions (2.2) together with (2.12), (2.17) 
lead immediately to the estimate:
$$
|k_{jl}(t)|\leq
W({\Bbb M}){\Bbb M}_0(t)
\| e^{-c|x-b_j(t)|}\chi(t)\|_2\leq W({\Bbb M}){\Bbb M}_0(t) 
{\Bbb M}_1(t)\rho^{\mu_1}(t).\tag 2.19$$
\subhead
2.4. Linear estimates
\endsubhead
To study
the behavior of solutions of the integral equation (2.10)
we need some estimates of the evolution operators 
${\Cal U}_{m}^A(t,\tau)P_{m}^A(\tau)$. The necessary 
estimates are collected 
in this subsection, the proofs being removed 
to the appendices.
\proclaim{Lemma 2.1} For any $x_0,x_1 \in {\Bbb R}^d$, 
$0\leq\tau\leq t\leq t_1$,
$$\|\left<x-x_0\right >^{-\nu_0 } {\Cal U}_j^A(t,\tau) P_j^A(\tau)f\|_2
\leq W(\hat {\Bbb M})
 \left<t-\tau\right>^{-d/2}\|\left<x-x_1\right >^{\nu_0} f\|_2.\tag 2.20$$
The function $W$ here is independent  of $x_0,x_1$
and $t_1$.
\endproclaim
See appendix 2 for the proof.
\par {\it Remark}. Due to the representation
$$ {\Cal U}_j^A(t,\tau) P_j^A(\tau)f=P_j^A(t)
{\Cal U}_0(t,\tau)f-
i\int_\tau^t ds {\Cal U}_j^A(t,s) P_j^A(s) 
(\tilde{\Cal V}_j(s)+R_j(s)){\Cal U}_0(s,\tau)f,$$
and the estimate
$$|(R_j(t) f)(x)|
\leq W({\Bbb M})e^{-c|x-b_{j}(t)|}(|\theta_j^\prime|+
|a_j^\prime|)\|e^{-c|x-b_{j}(t)|}f\|_2
$$
$$\leq W({\Bbb M})
e^{-c|x-b_{j}(t)|}{\Bbb M}_0(t)\|e^{-c|x-b_{j}(t)|}f\|_2,
\tag 2.21$$
(2.20) leads immediately to the inequality
$$\|\left<x-x_0\right >^{-\nu_0 } {\Cal U}_j^A(t,\tau) P_j^A(\tau)f\|_2
\leq W(\hat {\Bbb M}){(\|f\|_{p_1^\prime}+\|f\|_{p_2^\prime})
\over |t-\tau|^{d({1\over 2}-{1\over p_1})}
\left<t-\tau\right>^{d({1\over p_1}-{1\over p_2})}},\tag 2.22$$
where $2\leq p_1< {2d\over d-2}<p_2\leq\infty$,
${1\over p_i}+{1\over p_i^\prime}=1$, $i=1,2$.
Obviously, the same estimate is valid for $K_j(t,\tau)$:
$$\|\left<x-x_0\right >^{-\nu_0 }K_j(t,\tau)f\|_2
\leq W(\hat {\Bbb M}){(\|f\|_{p_1^\prime}+\|f\|_{p_2^\prime})
\over |t-\tau|^{d({1\over 2}-{1\over p_1})}
\left<t-\tau\right>^{-d({1\over p_1}-{1\over p_2})}}.\tag 2.23$$
 \par The key point of our analysis is the following lemma
that is essentially lemma 3.6 of [15].
\proclaim{Lemma 2.2}  
Introduce the operators
$T_{jki}(t,\tau)$, $j,k,i=1,\dots,N$, $i\neq k$ 
$$ T_{jki}(t,\tau)=A_j(t)K_k(t,\tau)A_i(\tau),$$
where $A_j(t)$ is the multiplication by $<x-b_j(t)>^{-\nu}$.
Then, for $0\leq t\leq t_1$
$$\int_0^t d\tau  \|T_{jki}(t,\tau)\|
\leq W(\hat {\Bbb M})(\epsilon_{ik}^{\nu_1}+{\Bbb M_0}(t)),$$
with some $\nu_1>0$. The norm $\|\cdot \|$ here
stands for the $L_2\rightarrow L_2$ operator norm.
\endproclaim
See appendix 3 for the proof.
\subhead
2.5. Estimates of the nonlinear terms
\endsubhead
Here we derive the necessary estimates of $D$, $D_j$.
We write $D$ as the sum:
$$D=D^0+D^1+D^2,$$
where
$$D^0=N_{00}+\sum_j \left(({\Cal V}_j-\tilde{\Cal V}_j)\vec\chi
+\tilde{\Cal V}_j\vec k_j+
e^{i\Phi_j\s_3}l(\s_j)\vec\xi_0(\cdot-b_j,E_j)\right),$$
$$N_{00}=F(|\psi_s|^2)
{\psi_s\choose
-\bar \psi_s}-\sum_j F(|w_j|^2){w_j\choose-\bar w_j}+
{\Cal V}(\psi_s)\vec\chi-\sum_j{\Cal V}_j\vec\chi,
$$
$$D^1=F(|\psi_s+\chi|^2)
{\psi_s+\chi\choose
-\bar \psi_s-\bar \chi}-F(|\psi_s|^2){\psi_s\choose-\bar \psi_s}-
{\Cal V}(\psi_s)\vec\chi-F(|\chi|^2){\chi\choose
-\bar \chi},$$
$$D^2=F(|\chi|^2){\chi\choose
-\bar \chi}.$$
In a similar way,
$$D_j=D_j^0+D^1+D^2,\quad j=1,\dots N,$$
where
$$D_j^0=N_{00}+ ({\Cal V}_j-\tilde{\Cal V}_j)\vec\chi-R_j\vec\chi+
\sum_k e^{i\Phi_k\s_3}l(\s_k)\vec\xi_0(\cdot-b_k,E_k).$$
Estimating $N_{00}$ by
$$|N_{00}|\leq c(1+|\chi|)
\sum\limits_{{j,k\atop k\neq j}}|w_j||w_k|,\tag 2.24$$
and using (2.12), (2.17), (2.19), (2.21)
one gets 
$$|D^0|, |D_j^0|\leq W({\Bbb M})
[(1+|\chi|)\sum\limits_{{i,k\atop i\neq k}}
e^{-c(|x-b_{i}|+|x-b_k|)}$$
$$+\sum\limits_ie^{-c|x-b_{i}|}
(|\l_i|+{\Bbb M}_0(t)|\chi|+ {\Bbb M}_0(t)M_1(t))].$$

which together with (2.16) leads to the inequality
$$\|D^0\|_{L_1\cap L_2},\,\|D^0_j\|_{L_1\cap L_2}
\leq
W({\Bbb M})[e^{-c|b_{jk}(t)|}+({\Bbb M}_0{\Bbb M}_1
+{\Bbb M}_1^2+{\Bbb M}_2^{2})\rho^{\mu_1}(t)].
\tag 2.25$$

\par Consider $D^1$,  $D^2$.  We estimate them as follows.
$$ |D^1+D^2|\leq W({\Bbb M})[|\psi_s||\chi|^2+|\chi|^3+|\chi|^{2p+1}],
\quad \roman{if}\,\,\,d=3,$$
$$|D^1|\leq W({\Bbb M})|\psi_s||\chi|^2,\,\, |D^2|\leq |\chi|^{2p+1},
\quad \roman{if}\,\,\,{1\over 2} < p <1,\tag 2.26$$
$$|D^1+D^2|\leq W({\Bbb M})|\chi|^{2p+1},
\quad \roman{if} \,\,\,p\leq {1\over 2}.$$

These inequalities imply for $r^\prime={2\over 1+p}$,
$$\|D^1+D^2\|_{L_1\cap L_{m^\prime}}
\leq W({\Bbb M})[{\Bbb M}_1^2+
{\Bbb M}_1^{2-{1\over p}}{\Bbb M}_2^{{1\over p}} +
{\Bbb M}_2^{1+{1\over p}}]\rho^{\mu_1}(t),\quad \roman{if}\,\,\, d=3,$$
$$\|D^1\|_{L_1\cap L_{m^\prime}}
+\|D^2\|_{L_{r^\prime}\cap L_{m^\prime}}
\leq W({\Bbb M})[{\Bbb M}_1^2+
{\Bbb M}_1^{2-{1\over p}}{\Bbb M}_2^{{1\over p}} +
{\Bbb M}_2^{1+p}]\rho^{\mu_1}(t),\,\,\, 
\roman{if}\,\,\,{1\over 2} < p <1 \tag 2.27$$
$$\|D^1+D^2\|_{L_{r^\prime}\cap L_{m^\prime}}\leq W({\Bbb M})
{\Bbb M}_2^{1+p}\rho^{\mu_1}(t),\quad \roman{if}\,\, \, 
p\leq{1\over 2}.$$

\subhead
2.6. Estimates of $\chi$ in $L_{2,loc }$
\endsubhead
To estimate $M_1(t)$ we use  representation (2.10). By (2.22),
for the first term $\roman{(I)}$ one has
$$
\|<y_j>^{-\nu}\roman{(I)}\|_2
\leq W({\Bbb M}){\Cal N}<t>^{-d/2}.\tag 2.28$$
\par Consider expression $\roman{(II)}$:
$$\|<y_j>^{-\nu}\roman{(II)}\|_2\leq W({\Bbb M}) {\Bbb M}_1(t)
\sum\limits_{{k,i\atop k\neq i}}\int_0^t ds \|T_{jki}(t,s)\|\rho^{\mu_1}(s).
\tag 2.29$$
By lemma 2.1,
$$\|T_{jki}(t,s)\|\leq W({\Bbb M})<t-s>^{-d/2}.$$
So, the integral in the right hand side of (2.29) can be estimated as 
follows.
$$\int_0^t ds \|T_{jki}(t,s)\|\rho^{\mu_1}(s)
\leq 
\left(\int_0^t ds 
\|T_{jki}(t,s)\|\rho^{d/2  }(s)\right)^{{2\mu_1\over d}}
\left(\int_0^t ds
\|T_{jki}(t,s)\|\right)^{1-{2\mu_1\over d}}$$
$$\leq W({\Bbb M})({\Bbb M}_0^\theta +\epsilon_{ik}^{\nu_2})
\left(\int_0^t ds <t-s>^{-d/2}
\rho^{d/2  }(s)\right)^{{2\mu_1\over d}}$$
$$\leq W({\Bbb M})({\Bbb M}_0^\theta +\epsilon_{ik}^{\nu_2})
\rho^{\mu_1}(t),$$
 $0<\theta=1-{2\mu_1\over d}$, $\nu_2=\theta\nu_1$. 
At the second step here  we  have  used lemma 2.2. Thus,
$$\|<y_j>^{-\nu}\roman{(II)}\|_2\leq W(\hat{\Bbb M})
({\Bbb M}_0^\theta +\epsilon_{ik}^{\nu_2}){\Bbb M}_1(t)
\rho^{\mu_1}(t).\tag 2.30$$
\par Consider the two last terms in the r.h.s. of (2.10). By (2.25),
(2.23) (with $p_1=2,\,p_2=\infty$)
one has
$$\|<y_j>^{-\nu}{\Cal U}_0(t,s)D^0(s)\|_2,
\,\|<y_j>^{-\nu}K_m(t,s)D_m^0(s)\|_2
\leq  W({\Bbb M})<t-s>^{-d/2}
$$
$$\times [\sum\limits_{{i,k\atop i\neq k}}
e^{-|b_{ik}(s)|}+
({\Bbb M}_0(t){\Bbb M}_1(t)+{\Bbb M}_1^2(t)+{\Bbb M}_2^{2}(t))
\rho^{\mu_1}(s)].\tag 2.31
$$
Using (2.27), (2.23) one can estimate the contribution of $D^1$, $D^2$
as follows.
$$\|<y_j>^{-\nu}{\Cal U}_0(t,s)(D^1(s)+D^2(s))\|_2,
\,\|<y_j>^{-\nu}K_m(t,s)(D^1(s)+D^2(s))\|_2$$
$$\leq  W({\Bbb M})[{\Bbb M}_1^2(t)+{\Bbb M}_2^{r_1}(t)]|t-s|^{-\mu_2}
<t-s>^{-\mu_1+\mu_2}\rho^{\mu_1}(s).\tag 2.32$$
Here $1<r_1=1+\min\{p,p^{-1}\}< 2$.
Combining (2.31), (2.32) and integrating with respect to $s$ one gets
$$\|<y_j>^{-\nu}\roman{(III)}\|_2, \|<y_j>^{-\nu}\roman{(IV)}\|_2
\leq W({\Bbb M})[\sum\limits_{{i,k\atop i\neq k}}
\int_0^t ds {e^{-|b_{ik}(s)|}\over
<t-s>^{d/2}}$$
$$+({\Bbb M}_0{\Bbb M}_1+{\Bbb M}_1^2+
{\Bbb M}_2^{r_1})\rho^{\mu_1}(t)],$$
or taking into account (2.14), (2.15),
$$\|<y_j>^{-\nu}\roman{(III)}\|_2, \|<y_j>^{-\nu}\roman{(IV)}\|_2
\leq W(\hat{\Bbb M})[\epsilon
+{\Bbb M}_0{\Bbb M}_1+
{\Bbb M}_1^2+{\Bbb M}_2^{r_1}]\rho^{\mu_1}(t).\tag 2.33$$
\par Combining (2.28), (2.30), (2.33), one obtains
$${\Bbb M}_1\leq W(\hat{\Bbb M})[{\Cal N}+
\epsilon^{\nu_2}+{\Bbb M}_0^\theta{\Bbb M}_1+
{\Bbb M}_1^2+{\Bbb M}_2^{r_1}
].$$
Changing if necessary the coefficient function 
$W$ one can simplify this inequality:
$${\Bbb M}_1\leq W(\hat{\Bbb M})[{\Cal N}+
\epsilon^{\nu_2}+{\Bbb M}_2^{r_1}].\tag 2.34$$

\subhead
2.7. Closing of the estimates
\endsubhead
Here we derive a 
$L_m$ estimate of $\chi$ which will close the system of the inequalities
for the majorants. 
To estimate $L_m$ - norm of $\chi$ we use representation (2.7).
By (2.24), (2.27),
$$\|N\|_{m^\prime}\leq W({\Bbb M})[\sum\limits_{{k,i\atop k\neq i}}
e^{-c|b_{ik}(t)|}+{\Bbb M}_1^2+{\Bbb M}_2^{r_1})
\rho^{\mu_1}(t)].$$
As a consequence,
$${\Bbb M_2}\leq W(\hat{\Bbb M})[{\Cal N}+\epsilon^{1-\mu_2}+
{\Bbb M}_1].\tag 2.35$$
Here we have made use of the inequality
$$\int_0^t ds {e^{-c|b_{ik}(s)|}\over
|t-s|^{\mu_2}}\leq  W(\hat{\Bbb M})
{\epsilon^{1-\mu_2}_{ik}\over <t-t_{ik}>^{\mu_2}},$$
which is an immediate consequence of (2.14), (2.15).

\par Combining  (2.18), (2.34), (2.35) one gets
$$
\hat{\Bbb M}_1,\,\hat{\Bbb M}_2\leq 
W(\hat{\Bbb M}) ({\Cal N}+\epsilon^{\nu_3}),\quad
\hat {\Bbb M}_0\leq W(\hat{\Bbb M})
({\Cal N}^2+\epsilon^{2\nu_3}),\tag 2.36$$
$\nu_3=\min\{{1\over 2},\nu_2,1-\mu_2\}>0$, 
the coefficient functions $W({\Bbb M})$ 
being independent 
of $t_1$.
These inequalities mean  
that for ${\Cal N}$ and $\epsilon$ sufficiently small
${\Bbb M}$ can belong either to a small
neighborhood of zero or
to some domain whose distance from zero is bounded from below
uniformly
with respect to ${\Cal N}$, $\epsilon$. 
Since $\hat{\Bbb M}_l$ are continuous functions of $t_1$ and for $t_1=0$
are small only the first possibility can be realized.
This means that for ${\Cal N}$ and $\epsilon$
in some finite vicinity of zero,
$$ {\Bbb M}_1(t),\, {\Bbb M}_2(t)\leq c({\Cal N}+\epsilon^{\nu_3}),\quad
 {\Bbb M}_0(t)\leq c({\Cal N}^2+\epsilon^{2\nu_3}),\quad 
0\leq t\leq t_1.$$
The constant $c$ here 
is independent of ${\Cal N}$, $\epsilon$, $t_1$. Since $t_1$ is
arbitrary these estimates are valid, in fact, for all $t\geq 0$.
More precisely, one has

$$ M_0(t)\leq c({\Cal N}^2+\epsilon^{2\nu_3}),\,\,\,
M_1(t)\leq c({\Cal N}+\epsilon^{\nu_3})\rho_\infty^{\mu_1}(t),
\,\,\,
M_2(t)\leq c({\Cal N}+\epsilon^{\nu_3})\rho_\infty^{\mu_2}(t),
\tag 2.37$$
where $\rho_\infty(t)$ is the weight function corresponding
to $t_1=\infty$:
$$\rho_\infty(t)=<t>^{-1} 
+\sum\limits_{{j, k\atop j\neq k}}<t-t_{jk}^\infty>^{-1},$$
$t_{jk}^\infty=0$ if $t_{jk}^0\leq 0$, and
$$\int_0^{t_{jk}^\infty}ds{v_{jk}(s)\cdot v_{jk}^0\over|v_{jk}^0|^2}
=t_{jk}^0,$$
if $t_{jk}^0> 0$.
\par By (2.15), (2.16), the estimates (2.36) imply
the existence of the limit trajectories
$\s_{+j}(t)=(\b_{+j}(t),E_{+j},b_{+j}(t), v_{+j})$, $j=1,\dots,N$,
$$b_{+j}(t)=v_{+j}t+b_{+j},\quad v_{+j}=v_{0j}+
\int_0^\infty dsv_{j}^\prime(s),$$
$$b_{+j}=b_{0j}+\int_0^\infty ds(c_{j}^\prime(s)+v_j(s)-v_{+j}),$$
$$\b_{+j}(t)=(E_{+j}-{|v_{+j}|^2\over 4})t+\b_{+j},\quad
E_{+j}=E_{0j}+
\int_0^\infty dsE_{j}^\prime(s),$$
$$\b_{+j}=\b_{0j}+
\int\limits_0^\infty ds\big(E_j-E_{+j}+{|v_j-v_{+j}|^2\over 4}+
\g_j^\prime-{1\over 2}v_j^\prime\cdot c_j\big).$$
Obviously, as $t\rightarrow +\infty$,
$$|E_j(t)-E_{+j}|,\, |v_{j}(t)-v_{+j}|=0(t^{-2\mu_1+1}),$$
$$|b_j(t)-b_{+j}(t)|,\,|\b_j(t)-
\b_{+j}(t)|=O(t^{-2\mu_1+2}).$$

\head
Appendix 1
\endhead
Here we outline the arguments needed for the proof of
the existence
of a decomposition (2.1) satisfying (2.2) for all $t\geq 0$. 
We begin with the following lemma.
\newline
Given $N$ solitons $w(\s_{0j})$, $\s_{0j}=(\b_{0j},E_{0j},b_{0j},v_{0j})$,
$j=1,\dots, N$, we define the effective coupling parameter
$\delta(\vec\s_0)$, $\vec \s_0=(\s_{01},\dots,\s_{0N})$,
$$\delta(\vec \s_0)=\max\limits_{j\neq k}(|v^0_{jk}|+|b^0_{jk}|)^{-1}.$$
For $\chi\in L_p({\Bbb R}^d)$,
$\vec \s=(\s_1,\dots,\s_N)$,
$\s_j=(\b_j,E_j,b_j,v_j)\in
{\Bbb R}\times {\Cal A}
\times {\Bbb R}^d\times {\Bbb R}^d$,
$j=1,\dots, N$, consider the functionals $F_{j,l}(\vec\s,\chi;\vec\s_0)$,
$j=1,\dots,N$, $l=0,\dots 2d+1$,
$$F_{j,l}(\vec\s,\chi;\vec\s_0)
=\left<\vec\chi+\sum\limits_{k=1}^N
\vec w(\s_{0k})-\vec w(\s_{k}),
\s_3\vec\zeta_l(\s_j)\right>,$$
where
$$\vec w={w\choose \bar w},\quad \vec\zeta_l(x,\s)=
e^{i\b\s_3+i{x\cdot v\over 2}\s_3}\vec\xi_l(x-b,E),\,\,\,
\s=(\b,E,b,v).$$
Set
$F_j=(F_{j,0},\dots, F_{j,2d+1})$,
$F=(F_{1},\dots, F_{N})$.
\proclaim
{Lemma A1.1} Let $E_{0j}\in {\Cal A}_0$, $j=1,\dots, N$.
There exist constants $n_0>0$, 
$\delta_0>0$, $K>0$, depending only on 
$E_{0j}$, $j=1,\dots, N$
such that if 
$\delta(\vec \s_0)\leq \delta_0$ and $\|\chi\|_p\leq n_0$
 then 
the equation 
$$F(\vec \s,\chi;\vec \s_0)=0$$
has a unique solution $\vec \s=(\s_1,\dots, \s_N)$,
$\vec\s$ being a $C^1$ function of $\chi$,
that satisfies
$$|\b_j-b_{0j}+{1\over 2}(v_j-v_{0j})\cdot b_{0j}|+
|E_{j}-E_{0j}|+|b_{j}-b_{0j}|+|v_{j}-v_{0j}|\leq K\|\chi\|_p.
\tag A1.1$$
\endproclaim
{\it Remark}. It follows directly from (A1.1)
that
\item{(i)} for some constant $K_1$
$$\|\chi+\sum\limits_{k=1}^N
 w(\s_{0k})- w(\s_{k})\|_p\leq K_1\|\chi\|_p,$$
\item{(ii)} if for some pair $(j,k)$, 
$t_{jk}^0\leq \kappa_0<r_{jk}^0>\,$
then the new collision time $t_{jk}=-{b_{jk}\cdot v_{jk}\over |v_{jk}|^2}$
satisfies a similar estimate with a constant 
$\kappa=\kappa_0(1+O(\|\chi\|_p)$. 
\par {\it Proof of Lemma }A1.1.
Let us pass from $\vec \s$ to a new system of parameters
$\vec \l=(\l_1,\dots, \l_N)$,
$$\l_j=(\b_j-\b_{0j}+{1\over 2}(v_j-v_{0j})\cdot b_{0j},
E_j,b_{j}-b_{0j},v_{j}-v_{0j}).$$
We represent $F(\vec \s,\chi;\vec \s_0)$ as the sum
$$F=F^0+F^1+F^2,$$
$F^0_{j}=\Phi(\l_j,E_{0j})$, $\Phi=(\Phi_0,\dots \Phi_{2d+1})$,
$$\Phi_l(\l,E)=\left<\vec\xi_0(E)-\vec\zeta_0(\l),\vec\zeta_l(\l)\right>,$$
$$F^1_{j,l}=\sum\limits_{k,\,k\neq j}
\left<\vec\zeta_0(\s_{0k})-
\vec\zeta_0(\s_{k}),\vec\zeta_l(\s_{j})\right>.$$
At last,
$$ F^2_{j,l}=G_l(\l_j,f_j),\quad 
\chi(x)=e^{i\b_{0j}+i{v_{0j}\cdot x\over 2}}f_j(x-b_{0j}),$$
$G_l(\l,f)=<\vec f,\s_3\vec\zeta_l(\l)>$ is a $C^1$ function of 
$f$ and $\l$.
\par The direct calculations give
$$|\det \nabla_\l \Phi(\l,E)|\big |_{\l=(0,E,0,0)}
=e^2(E)n^{2d}(E).\tag A1.2$$
Set $\vec \l_0=(\l_{01},\dots,\l_{0N}),\quad
\l_{0j}=(0,E_{0j},0,0).$ By (A1.2),
$$ |\det \nabla_{\vec \l}F^0| \big |_{\vec \l=\vec\l_0}
=\prod\limits_{j=1}^Ne^2(E_{0j})n^{2d}(E_{0j})\tag A1.3$$
is nonzero if $E_{0j}\in {\Cal A}_0$, $j=1,\dots, N$.
\par Consider $F^1$. It is not difficult to check that
for $\vec\l$ in some finite vicinity of $\vec \l_0$
the derivative $\nabla_{\vec \l}F^1$ satisfies
the inequality
$$|\nabla_{\vec \l}F^1|\leq C\delta(\vec\s_0),\tag A1.4$$
constant $C$ depending only on $E_{0j}$.
\par By the implicit  function theorem,
the desired result is a direct consequence of
(A1.3), (A1.4). \hfill{$\blacksquare$}
\par To prove the existence of a decomposition (2.1)
satisfying (2.2) for all $t>0$ we use some standard continuity type
arguments. Since $\psi \in C({\Bbb R}\rightarrow H^1)$
there exists a small interval $[0,t_1]$
where the constructions of lemma A1.1 can be used.
This leads to a representation  (2.1)
satisfying the orthogonality conditions for $t\in [0,t_1]$.
For the components of such a representation
estimates (2.15), (2.36) give
$$|E_{0j}-E|\leq  C ({\Cal N}^2+\epsilon^{2\nu_3}),\quad 
(|v_{jk}|+|b_{jk}|)^{-1}\leq C\epsilon,$$
$$\|\chi(t)\|_m\leq C ({\Cal N}+\epsilon^{\nu_3}),$$
which allows us to extend decomposition (2.1), (2.2)
on a larger interval $[0,t_1+t_2]$
with some $t_2>0$. On this new interval the same estimates 
hold, so one can continue the procedure with 
 steps of the same length $t_2$. As a result, one gets a decomposition
(2.1) satisfying (2.2) for all $t\geq 0$.

\head
Appendix 2
\endhead
Here we prove lemma 2.1.
Consider the equation
$$i\chi_t={\Cal L}(t)\chi,\quad
{\Cal L}(t)=(-\triangle+E)\s_3+{\Cal V}(t)+i[ P^\prime(t),P(t)],
\tag A2.1$$
$$
{\Cal V}(t)=T(t)V(E)T^*(t), \quad
P(t)=T(t)\hat P(E)T^*(t),$$
where $T(t)=B_{\theta(t),a(t),0}$.
We denote the corresponding propagator by $U(t,\tau)$.
Clearly,
$${\Cal U}_j^A(t,\tau)=B_{\s_{0j}(t)}
U(t,\tau)\big|_{\theta=\theta_j,\,a=a_j,E=E_{0j}}B_{\s_{0j}(\tau)}.$$
We shall assume that for some positive constants $n$, $R$, $\delta_1$,
$$|\theta^\prime(t)|+|a^\prime(t)|\leq n,\tag A2.2$$
$$
|\theta^{\prime\prime}(t)|+|a^{\prime\prime}(t)|
\leq \sum\limits_{l=0}^L <R(t-t_l)>^{-2-\delta_1},\tag A2.3$$
$t\in {\Bbb R}_+$.
Here $L\in {\Bbb N} $, $0=t_0<t_1<\dots<t_L$. One 
has the following lemma.
\proclaim{Lemma A2.1}
For any $x_0,x_1 \in {\Bbb R}^d$, $t\geq 0$,$\tau\geq 0$,
$$\|\left<x-b_0\right >^{-\nu_0 } U(t,\tau) P(\tau)f\|_2
\leq C \left<t-\tau\right>^{-d/2}\|\left<x-x_1\right >^{\nu_0} f\|_2,
$$
provided n is sufficiently small and $R$ is sufficiently large:
$n+R^{-1}\leq C$.
\endproclaim
In this appendix we use $C$
as a general notation for constants that depend only on
$M,\delta, E$ and can be chosen uniformly 
with respect to $E$ in compact subsets of ${\Cal A}_0$.
\par It follows from (2.12), (2.14), (2.15), (2.16) that
for $\hat {\Bbb M}$ in some finite vicinity of zero
  the functions $\theta_j$, $a_j$ satisfy 
assumptions (A2.2), (A2.3)
with $\delta_1=2\mu_1-2$, $t_l$, $l=1,\dots, L$, being 
the collision times 
$t_{ik}$, $i,k=1,\dots N,\,\, i\neq k$. So, lemma A2.1 implies lemma 2.1.
\par {\it Proof of lemma} A2.1.
Lemma A2.1 follows from proposition 1.1 by a 
simple perturbation argument.
 On the intervals
$[t_l, t_{l+1}]$, $l=0,\dots L-1$ we introduce the following 
linear approximations
$\theta^{l}(t)$, $a^l(t)$ of 
 $\theta(t)$, $a(t)$:
$$\theta^{l}(t)=\theta(t)-\int_{t_l}^tds\int_{t_l}^s
ds_1\left(1-\eta({s_1-t_l\over t_{l+1}-t_l})\right)
\theta^{\prime\prime}(s_1)$$
$$-
\int_{t}^{t_{l+1}}ds\int^{t_{l+1}}_s
ds_1\eta({s_1-t_l\over t_{l+1}-t_l})\theta^{\prime\prime}(s_1),$$
$$a^{l}(t)=a(t)-\int_{t_l}^tds\int_{t_l}^s
ds_1\left(1-\eta({s_1-t_l\over t_{l+1}-t_l})\right)a^{\prime\prime}(s_1)$$
$$-
\int_{t}^{t_{l+1}}ds\int^{t_{l+1}}_s
ds_1\eta({s_1-t_l\over t_{l+1}-t_l})a^{\prime\prime}(s_1).$$
Here $\eta\in C^\infty({\Bbb R})$, $\eta(\xi)=\cases 
1\,\,\roman {for} |\xi|\leq {1\over 4},\cr
0\,\,\roman {for} |\xi|\geq {3\over 4}.\cr\endcases$
\newline
For $t\in [t_L,\infty)$ we define the corresponding
$\theta^{L+1}(t)$, $a^{L+1}(t)$
as follows.
$$\theta^{L+1}(t)=\theta(t)-\int^\infty_tds\int^\infty_s
ds_1\theta^{\prime\prime}(s_1),$$
$$a^{L+1}(t)=a(t)-\int^\infty_tds\int^\infty_s
ds_1a^{\prime\prime}(s_1).$$
Clearly, for $t\in [t_l,t_{l+1}]$, $l=0,\dots, L$,
$t_{L+1}=\infty$, one has
$$|\theta(t)-\theta^l(t)|,\,\,|a(t)-a^l(t)|
\leq CR^{-2},\tag A2.4$$
$$|{d\theta^l\over dt}|,\,
|{da^l\over dt}|\leq C(n+R^{-1}).
\tag A2.5$$
\par On the interval $[t_l,t_{l+1}]$
one can pick out the leading term of (A2.1)
in the form
$$i\chi_t={\Cal L}^l(t)\chi,\quad {\Cal L}^l(t)=
(-\triangle +E)\s_3 +{\Cal V}^l(t),\tag A2.6$$
$$ {\Cal V}^l(t)=
T^l(t)V(E^{l})T^{l^*},\quad
T^l(t)=B_{\triangle^l(t),a^l(t),r^l},$$
$$
\triangle^l(t)=\theta^l(t)-{r^l\cdot a^l(t)\over 2},\,\,
r^l={da^l\over dt},\,\, E^l=E+{d\theta^l\over dt}
-{|r^l|^2\over 4}.$$
We denote the propagator corresponding to (A2.6)
by $U^l(t,\tau)$.
Clearly, 
$$U^l(t,\tau)=T^l(t)e^{-i(t-\tau)L(E^l)}T^{l^*}(\tau),\quad
P^l(t)U^l(t,\tau)=U^l(t,\tau)P^l(\tau),$$
where
$P^l(t)=T^l(t)\hat P(E^{l})T^{l^*}(t)$.
\par Consider the 
expression $\chi(t)\equiv U(t,\tau)P(\tau)f$,
$t_l\leq \tau <t_{l+1}$.

For $t_l\leq t\leq t_{l+1}$ 
we write  $\chi(t)$  as 
the sum $\chi=h+k$, $h(t)=P^l(t)\chi(t)$.
Since $\chi(t)=P(t)\chi(t)$, the $2d+2$ dimensional 
component $k$
is controlled by $h$:
$$\|e^{\g|x-a(t)|}k(t)\|_2\leq
C(|\theta(t)-\theta^l(t)|+|a(t)-a^l(t)|+|r^l|+|E-E^l|)
\| e^{-\g|x-a(t)|}\chi(t)\|_2$$
$$\leq C (R^{-1}+n)\| e^{-\g|x-a(t)|}\chi(t)\|_2\tag A2.7$$
for some $\g>0$,
provided $n$, $R^{-1}$ are sufficiently small.
In the last inequality  we used  (A2.4), (A2.5).
\par For $h$ one  can write the following integral representation
$$h(t)=P^l(t)h_0(t)-i\int_\tau^tds P^l(t)U^l(t,s)[{\Cal V}^l(s)h_0(s)
+R^l(s)\chi(s)],\tag A2.8$$
where
$$h_0(t)=e^{i(\triangle -E)(t-\tau)\s_3}P(\tau)f,$$
$$R^l(t)={\Cal V}(t)-{\Cal V}^l(t)
+i[P^\prime(t),P(t)].$$
Obviously,
$$|{\Cal V}(x,t)-{\Cal V}^l(x,t)|\leq C
|\theta(t)-\theta^l(t)|+|a(t)-a^l(t)|+|r^l|+|E-E^l|)
 e^{-\g|x-a(t)|}$$
$$\leq C (R^{-1}+n) e^{-\g|x-a(t)|},\tag A2.9$$
$$|[P^\prime(t),P(t)]f|
\leq Cne^{-\g|x-a(t)|}\| e^{-\g|x-a(t)|}f\|_2.\tag A2.10$$
Estimates (A2.7), (A2.9), (A2.10) and representation (A2.8) 
together with proposition 1.1
imply immediately that for $t_l\leq \tau\leq t\leq t_{l+1}$
and for any $\xi\in {\Bbb R}$ the following inequality holds
$$\|\left<x-a(t)\right>^{-\nu_0}\chi(t)\|_2<t-\tau+\xi>^{d/2}$$
$$\leq C \sup\limits_{\tau\leq s\leq t}
\left(\|\left<x-a(s)\right>^{-\nu_0}e^{it\triangle\s_3(s-\tau)}P(\tau)f\|_2
<s-\tau+\xi>^{d/2}\right),\tag A2.11$$
where $C$ do not depend on $\xi$.
(A2.11) implies in particular, that
$$\|\left<x-a(t)\right>^{-\nu_0}U(t,\tau)P(\tau)f\|_2\leq C <t-\tau>^{-d/2}
\|\left<x-x_0\right>^{\nu_0}f\|_2,\tag A2.12$$
$x_0\in {\Bbb R}^d$,
$t_l\leq \tau\leq t\leq t_{l+1}$, $l=0,\dots, L$.
\par To prove that this estimate is in fact true for
any $0\leq\tau\leq t$ we use the induction  arguments.
Assume that one has (A2.12) for $\tau\leq t\leq t_{l}<\infty$.
We need to show  that then the same 
is true for  $\tau\leq t_l<t\leq t_{l+1}$.
For $ t\in (t_{l},t_{l+1}]$ we write
$U(t,\tau)P(\tau)f=U(t,t_{l})U(t_l,\tau)P(\tau)f$.
Using (A2.12) and the representation 
$$
U(t,\tau)P(\tau)f=
e^{i(t-\tau)(\triangle-E)\s_3}P(\tau)f$$
$$-
i\int_\tau^t ds e^{i(t-s)(\triangle-E)\s_3}\left({\Cal V}(s)+
i[P^\prime(s),P(s)]\right)U(s,\tau)P(\tau)f,
\tag A2.13$$
one checks easily that
$$ \|<x-a(t)>^{-\nu_0}e^{i\triangle \s_3(t-t_{l})}U(t_l,\tau)P(\tau)f\|_2
\leq
C <t-\tau>^{-d/2} \|\left<x-x_0\right>^{\nu_0}f\|_2.$$
By (A2.11), this implies that (A2.12) is valid 
for $0\leq \tau\leq t\leq t_{l+1}$ and thus, for any 
$0\leq\tau\leq t$. Moreover, by (A2.13) one can replace $a(t)$
in the left hand side of (A2.12) by any $x_1\in {\Bbb R}^d$:
$$\|\left<x-x_1\right>^{-\nu_0}U(t,\tau)P(\tau)f\|_2\leq C
\|\left<x-x_0\right>^{\nu_0}f\|_2,\quad 0\leq\tau\leq t.
\qquad\quad \blacksquare$$

\head
Appendix 3
\endhead
Here we prove lemma 2.2.
We start by proving a similar result for the "free" operators
$T^0_{jkl}(t,\tau)$:
$$T^0_{jkl}(t,\tau)= A_j(t)
\int_\tau^t d\rho{\Cal U}_0(t,\rho)
\tilde{\Cal V}_{k}(\rho){\Cal U}_0(\rho,s) A_i(\tau).$$

\proclaim{Lemma A3.1}
For $i\neq k$,
$t\geq 0$, one has
$$\int_0^t d\tau  \|T^0_{jki}(t,\tau)\|
\leq W({\Bbb M})\epsilon_{ik}^{\nu_1}\tag A3.1$$
with some $\nu_1>0$.
\endproclaim
\par{\it Proof}.
Since 
$$\|T^0_{jkl}(t,\tau)\|\leq C<t-\tau>^{-d/2},$$
one has 
$$I(t)=\int_0^t d\tau  \|T^0_{jki}(t,\tau)\|\leq
C{t\over <t>}.\tag A3.2$$
In this appendix the constants $C$ depend only on $E_{0k}$.
\par For $t\geq 2\rho$, where $\rho$ is a small positive number,
we write the integral $I(t)$
as a sum of two terms $I(t)=I_0(t)+I_1(t)$,
$$I_0(t)=\int_{0}^{t-2\rho}d\tau\|T^{\rho}_{jki}(t,\tau)\|,$$
$$T^{\rho}_{jki}(t,\tau)= A_j(t)
\int_{\tau+\rho}^{t-\rho} ds
{\Cal U}_0(t,s)\tilde {\Cal V}_k(s){\Cal U}_0(s,\tau)
 A_i(\tau),$$
$I_1(t)$ being the rest.
Obviously,
$$I_1(t)\leq C\rho.\tag A3.3$$
\par Consider $I_0(t)$. To estimate this expression
we write $T^{\rho}_{jki}(t,\tau)$
in the form
$$T^{\rho}_{jki}=
\pmatrix {\Cal T}_{jki}^{11}&{\Cal T}_{jki}^{12}\cr
-{\Cal T}_{jki}^{21}&-{\Cal T}_{jki}^{22}\cr\endpmatrix,
$$
where
$${\Cal T}_{jki}^{11}(t,\tau)= A_j(t)
\int_{\tau+\rho}^{t-\rho} ds
e^{i(t-s)\triangle}{\Cal V}^1_k(s)e^{i(s-\tau)\triangle}
 A_i(\tau),$$
$${\Cal T}_{jki}^{12}(t,\tau)= A_j(t)
\int_{\tau+\rho}^{t-\rho} ds
e^{i(t-s)\triangle}{\Cal V}^2_k(s)e^{-i(s-\tau)\triangle}
 A_i(\tau),$$
$${\Cal V}^1_k(x,t)=V_1(x-\tilde b_{k},E_0),\quad
 {\Cal V}^2_k(x,t)=
e^{2i\tilde\Phi_k(x,t)}V_2(x-\tilde b_{k}(t),E_0),$$
$${\Cal T}_{jki}^{22}(t,\tau)f=
\overline{{\Cal T}_{jki}^{11}(t,\tau)\bar f},\quad 
{\Cal T}_{jki}^{21}(t,\tau)f=
\overline{{\Cal T}_{jki}^{12}(t,\tau)\bar f}
\tag A3.4$$
\par Consider ${\Cal T}_{jki}^{11}(t,\tau)$.
Since Hilbert-Schmidt norms dominate operator norms, we have
$$
\|{\Cal T}_{jki}^{11}(t,\tau)\|^2\leq C
\int\limits_{{\Bbb R}^{2d}} dxdy<x>^{-2\nu}<y>^{-2\nu}|
{\Cal B}_{jki}^1(t,\tau)|^2,$$
where
$${\Cal B}_{jki}^1(t,\tau)=\int_{\tau+\rho}^{t-\rho}ds
(t-s)^{-d/2} (s-\tau)^{-d/2}$$
$$\times
\int\limits_{{\Bbb R}^d} dz 
e^{{i|x-z+ b_j(t)-\tilde b_k(s)|\over 4(t-s)}}
V_1(z)
e^{{i|z-y+\tilde b_k(s)- b_i(\tau)|\over 4(s-\tau)}}.$$
Integrating by parts in the second integral and 
taking into account (2.13)
one gets immediately the estimate
$$|{\Cal B}_{jki}^1(t,\tau)|\leq W({\Bbb M})(<x>+<y>)\rho^{-1}$$
$$\times
\int\limits_{\tau+\rho}^{t-\rho}
 ds (t-s)^{-d/2} (s-\tau)^{-d/2}<d_{jki}(t,s,\tau)>^{-1},$$
where 
$$d_{jki}(t,s,\tau)={ \tilde b_{jk}(t)\over (t-s)}+
{ \tilde b_{ik}(\tau)\over (s-\tau)}.$$
Here the function $W$ do not depend on $\rho$.
As a consequence, one has  for 
$0\leq \a<\min\{1,{d\over 4}-{1\over 2}\}$, 
$|\tilde b_{jk}(t)|+|\tilde b_{ik}(\tau)|>0$,
$$\|{\Cal T}^{11}_{jki}(t,\tau)\|
\leq  W({\Bbb M}) \rho^{-1-d+2\alpha}
\int\limits_\tau^t ds <t-s>^{-d/2+\alpha}
<s-\tau>^{-d/2+\alpha}$$
$$\times
|\tilde b_{jk}(t)(s-\tau)+\tilde b_{ik}(\tau)(t-s)|^{-\a}
\leq  W({\Bbb M}) \rho^{-1-d+2\alpha}$$
$$\times
 <t-\tau>^{-d/2+2\alpha}
(|\tilde b_{jk}(t)-\tilde b_{ik}(\tau)|+
(t-\tau)|\tilde b_{jk}(t)|)^{-\a}.\tag A3.5$$
Here we made use of the obvious inequality
$$\int\limits_{\Bbb R}ds <s>^{-a}<s-\rho>^{-a}|d_1s+d_2|^{-\a}\leq C
<\rho>^{-a+\a}(|d_1|+|d_2|)^{-\a},$$
provided
 $a>1$,  $0\leq \a<1$, 
$d_1$, $d_2\in {\Bbb R}^{d}$, $C$ being independent of
$d_1$, $d_2$.
\par Integrating (A3.5) and taking into account (2.14,15) 
one gets finally,
$$\int_{0}^{t-2\rho}d\tau\|{\Cal T}^{11}_{jki}(t,\tau)\|
\leq  W({\Bbb M})\rho^{-1-d+2\alpha}
\epsilon_{ik}^\a.\tag A3.6$$

\par In a similar way, one has for ${\Cal T}^{12}_{jki}(t,\tau)$
$$
\|{\Cal T}_{jki}^{12}(t,\tau)\|^2\leq C
\int\limits_{{\Bbb R}^{2d}} dxdy<x>^{-2\nu}<y>^{-2\nu}|
{\Cal B}_{jki}^2(t,\tau)|^2,$$
$${\Cal B}_{jki}^2(t,\tau)=\int_{\tau+\rho}^{t-\rho}ds
(t-s)^{-d/2} (s-\tau)^{-d/2}$$
$$\times
\int\limits_{{\Bbb R}^d} dz 
e^{{i|x-z+ b_j(t)-\tilde b_k(s)|\over 4(t-s)}}
e^{2i\tilde \Phi(z+\tilde b_k(s),s)}
V_2(z)
e^{-{i|z-y+\tilde b_k(s)- b_i(\tau)|\over 4(s-\tau)}},$$
which implies
$$
\int_{0}^{t-2\rho}\|{\Cal T}^{12}_{jki}(t,\tau)\|
\leq  W({\Bbb M}) \rho^{-1-d+2\alpha}\int\limits_0^td\tau
\int\limits_\tau^t ds <t-s>^{-d/2+\alpha}$$
$$\times <s-\tau>^{-d/2+\alpha}
|\tilde b_{jk}(t)(s-\tau)-\tilde b_{ik}(\tau)(t-s)|^{-\a}$$
$$
\leq  W({\Bbb M})\rho^{-1-d+2\alpha}
\epsilon_{ik}^\a.\tag A3.7$$
Combining (A3.2), (A3.3), (A3.4), (A3.6), (A3.7)  one obtains
$$
I(t)\leq W({\Bbb M})(\rho +\rho^{-1-d+2\alpha}
\epsilon_{ik}^\a),$$
which leads immediately to (A3.1) 
with $\nu_1\leq {\a\over 2+d-2\a}$. 
\hfill $\blacksquare$
\par Let us introduce the operators $T^1_{jki}(t,\tau)$:
$$T^1_{jkl}(t,\tau)= A_j(t)
\int_\tau^t ds{\Cal U}_0(t,s)
\tilde{\Cal V}_{k}(s)
(I-P^A_k(s)){\Cal U}_0(s,\tau) A_i(\tau).$$
It is not difficult to check that for any $\a\leq 1$,
$$\|A_j(t){\Cal U}_0(t,s)
\tilde{\Cal V}_{k}(s)
(I-P^A_k(s)){\Cal U}_0(s,\tau) A_i(\tau)\|$$
$$\leq   
W({\Bbb M})<t-s>^{-d/2}<s-\tau>^{-d/2+\a}<\tilde b_{ik}(\tau)>^{-\a}.$$
As a consequence,
$$\int_0^t d\tau\|T^1_{jkl}(t,\tau)\|\leq  W({\Bbb M})
\int_0^t d\tau<t-\tau>^{-d/2+\a}<\tilde b_{ik}(\tau)>^{-\a}
\leq  W({\Bbb M})\epsilon_{ik}^\a.\tag A3.8$$
At the last step here we have used (2.14), (2.15).
\par {\it Proof of lemma 2.2}. This lemma follows directly from
(A3.1), (A3.8)  and the following representation
$$T_{jkl}(t,\tau)=T_{jki}^{0}(t,\tau)-T_{jki}^{1}(t,\tau)$$
$$-i\int_\tau^td\rho\int_\tau^\rho dsA_j(t){\Cal U}_0(t,\rho)
\tilde{\Cal V}_{k}(\rho)P^A_k(\rho){\Cal U}_k^A(\rho,s)
 R_k(s){\Cal U}_0(s,\tau) A_i(\tau)\tag A3.9$$
$$-i\int_\tau^t d\rho
A_j(t){\Cal U}_0(t,\rho)
\tilde{\Cal V}_{k}(\rho)P^A_k(\rho)A_k^{-1}(\rho)
T_{kki}^{0}(\rho,\tau) \tag A3.10$$
$$-\int_\tau^t d\rho\int_\tau^\rho dsA_j(t){\Cal U}_0(t,\rho)
\tilde{\Cal V}_{k}(\rho)P^A_k(\rho){\Cal U}_k^A(\rho,s)
[\tilde{\Cal V}_k(s)+ R_k(s)]
A_k^{-1}(s)T_{kki}^{0}(s,\tau).\tag A3.11$$
We estimate the right hand side of this representation term by term.
Using lemma 2.1 and inequality (2.21) one gets 
$$\int_0^t d\tau\|(A3.9)\|\leq W(\hat{\Bbb M}){\Bbb M}_0(t).\tag A3.12$$
Expression (A3.10) can be estimated as follows
$$ \int_0^t d\tau\|(A3.10)\|
\leq W({\Bbb M})\int_0^td\tau\int_\tau^t d\rho
<t-\rho>^{-d/2}\|T_{kki}^{0}(\rho,\tau)\|
\leq  W({\Bbb M})\epsilon_{ik}^\a.\tag A3.13$$
In a similar way,
$$\int_0^t d\tau\|(A3.11)\|
\leq W(\hat {\Bbb M})\int_0^td\tau\int_\tau^t d\rho
\int_\tau^\rho ds<t-\rho>^{-d/2}$$
$$<\rho-s>^{-d/2}
\|T_{kki}^{0}(s,\tau)\|
\leq  W(\hat{\Bbb M})\epsilon_{ik}^\a.\tag A3.14$$
Combining (A3.1), (A3.8), (A3.12), (A3.13), (A3.14) 
one gets lemma 2.2.
\hfill $\blacksquare$

\head
Appendix 4
\endhead
Here we discuss the proof of proposition 1.1.
Since only the weighted estimates are needed,
rather then follow [8, 37, 38] we use the approach of [16, 17, 18].
It turns out that
the arguments of [16, 17, 18] can be applied almost 
without modifications. So, 
we describe only the main steps of the proof,
referring the reader to [16, 17, 18]
for most of the details.
\par We start be recalling  briefly some basic properties 
of the free resolvent 
$R_0(\l)=\pmatrix (-\triangle+E-\l)^{-1}&0\cr
0&-(-\triangle+E+\l)^{-1}\cr\endpmatrix$.
Let  $H^{t,s}$ stand for the weighted Sobolev spaces:
$$H^{t,s}=\{f,\, \|f\|_{H^{t,s}}\equiv
\|<x>^s(1-\triangle)^{t/2}f\|_2<\infty\}.$$
We denote by $B(H^{s,t},H^{s_1,t_1})$
the space of bounded operators from $H^{s,t}$ to
$H^{s_1,t_1}$.
Set $L_2^s=H^{0,s}$, $B(H^{s,t})=B(H^{s,t},H^{s,t})$.
If $s>1$ and $t\in {\Bbb R}$ the resolvent $R_0(\l)$
which is originally defined as $B(L_2)$ valued analytic function of 
$\l\in{\Bbb C}\setminus (-\infty,-E]\cup [E,\infty)$ 
can be extended continuously to the $\overline{{\Bbb C}^+}=
\{\im \l\geq 0\}$ when considered as a $B(H^{s,t},H^{-s,t+2})$
valued function.
The following
properties of $R_0(\l)$ are well known, see [16, 17, 18, 37, 38]
and references therein.
\proclaim{Lemma A4.1} 
Let $k=0,1,\dots$. If $s>k+{1/2}$, then  the derivative
$R_0^{(k)}(\l)
\in B(H^{s,0},H^{-s,0})$ is continuous in $\l\in 
\overline{{\Bbb C}^+}\setminus\{E,-E\}$, with
$$ R_0^{(k)}(\l)=O(|\l|^{-(k+1)/2}),\tag A4.1$$
in this norm as $\l\rightarrow \infty$
in $\overline{{\Bbb C}^+}$.
\endproclaim
The behavior of $R_0(\l)$ for $\l$ close to $\pm E$
is described by the following lemma,
see again [16,17,18].
\proclaim{Lemma A4.2} As $\l\rightarrow  E$,
 $R_0(\l)$ admits the 
following asymptotic expansion in $B(H^{s,t},H^{-s,t+2})$.
\newline For $m$ odd: 
$$R_0(\l)=\sum\limits_{j=0}^lG_{j,0}( \lambda - E)^{j}
+\sum\limits_{j=0}^{l} G_{j,1}( \lambda - E)^{j+{1\over 2}}+
O((\lambda -E)^{l+1}),\tag A4.2$$
for $m$ even:
$$R_0(\l)=\sum\limits_{j=0}^lG_{j,0}( \lambda -E)^{j}+
\ln (\lambda -E)
\sum\limits_{j=0}^{l} G_{j,1}( \lambda -E)^{j}
+o((\lambda -E)^{l}),\tag A4.3$$
where  $l=0,1,\dots$, $s>C(l,d)$, 
the coefficients $G_{j,k}$ belong to $ B(H^{s,t},H^{-s,t+2})$,
$G_{j,1}=0$ for 
$j<{d-3\over 2}$ if $d$ is odd and for
$j< {d-2\over 2}$ if  $d$ is even.
Representations {\rm (A4.2), (A4.3)} can be differentiated with
respect to $\l$ any number of times. 
\endproclaim
Here $(\lambda -E)^{1/2}$ and $\ln (\lambda-E)$ 
are defined on the complex plane
with the cut along
$[E,\infty)$.
The explicit expressions for the constants $C(l,d)$
can be found in [17, 18, 19]. Similar expansions hold
as $\l\rightarrow -E$.
\par For $\l\in [E,\infty)$, consider the operator
$$ I+R_0(\l+i0)V: L_2^{-s}\rightarrow L_2^{-s},$$
$s>1$.
\proclaim{Lemma A4.3}
Let $E\in {\Cal A}_0$. Then $\Ker  (I+R_0(\l+i0)V)$
is trivial.
\endproclaim
\par {\it Proof}. We start by the case $\l=E$.
Let 
$\psi\in \Ker ( I+G_0 V)$. This implies that $\psi$ 
belongs to $L_2({\Bbb R}^d)+<x>^{-(d-2)}L_\infty({\Bbb R}^d)$ and satisfies
$$L\psi=E \psi.$$
Hypothesis H3 then allows us to conclude that $\psi=0$.
\par We consider next the case $\l>E$.
Let $\psi\in \Ker ( I+R_0(\l+i0) V)$. Since $V$ is spherically
symmetric, one can assume that
$\psi(x) =f(r)Y(\omega)$, $r=|x|$, $\omega ={x\over |x|}$,
$f\in L_2({\Bbb R}_+; r^{d-1}<r>^{-2s} dr)$
and $Y\in L_2(S^{d-1})$,
$$\triangle_{S^{d-1}} Y=\mu_n Y,\quad \mu_n=n(d-2+n),$$
for some $n\in\{0,1,\dots\}$.
 Then $f$ has to satisfy
$$l_nf\equiv 
\left[(-{d^2\over dr^2}-
{d-1\over r}{d\over dr}+E+{\mu_n\over r^2})\s_3 +V\right]f=\l f,
\tag A4.4$$
$$f^\prime(0)=0 \,\,\, \roman{if}\quad n=0,\quad
f(0)=0\,\,\,\roman{if}\quad n>0,$$
and as $r\rightarrow \infty$,
$$f=cr^{-{(d-2)\over 2}}H_{\nu}^{(1)}(k r){1\choose 0} +
O(e^{-\gamma r}),\quad
\g >0,\tag A4.5$$
for some constant $c$.
Here $k=(\l-E)^{1/2}>0$, $\nu=n+{(d-2)\over 2}$,
 $H_\nu^{(1)}$ is the first Hankel function.
Asymptotic representation (A4.5)
can be differentiated with respect to $r$ any number of times.
\par The Wronskian
$$w(f,g)=r^{d-1}(<f^\prime,g>_{{\Bbb R}^2}-<f,g^\prime>_{{\Bbb R}^2})$$
does not depend on $r$ if $f$ and $g$ are solutions of 
(A4.4). Calculating $w(f,\bar f)$ one gets
$$2ik|c|^2 =0,$$
which implies that $\psi \in L_2$. Since $E\in {\Cal A}_0$,
this means that $\psi=0$.\hfill $\blacksquare$

Consider the full resolvent
$R(\l)=(L-\l)^{-1}$.
$R(\l)\hat P$ ($R(\l)$) is a $B(L_2)$ valued holomorphic 
(meromorphic with the only pole in 
zero)
function of 
$\l\in {\Bbb C}\setminus (-\infty,-E]\cup [E,\infty)$.
$R(\l)$ satisfies the relations
$$\s_1R(\l)\s_1=-R(-\l).
\tag A4.6 $$
The analytic properties of $R(\l)$ near the cuts
$ (-\infty,-E]$, $ [E,\infty)$ are collected 
in the two following lemmas. 
In both of them we assume that $E\in{\Cal A}_0$.
\proclaim{Lemma A4.4} For $s>1$,
$R(\l)\hat P$ can be extended continuously 
to  $\overline{{\Bbb C}^+}$ as a $B(L_2^s,L_2^{-s})$
valued function. Moreover,
if $s>k+{1\over 2}$ then $R^{(k)}(\l)\hat P$ exists and continuous
for $\l\in \overline{{\Bbb C}^+}\setminus \{E,-E\}$
and 
$$R^{(k)}(\l)\hat P=O(|\l|^{-(k+1)/ 2})\tag A4.7$$
in $B(L_2^s,L_2^{-s})$ as $\l\rightarrow \infty $
in $ \overline{{\Bbb C}^+}$.
\endproclaim
\proclaim{Lemma A4.5}
As $\l\rightarrow  E$, $R(\l)$ admits the 
following asymptotic expansion in $B(L_2^s,L_2^{-s})$.
\newline For $m$ odd: 
$$R(\l)=\sum\limits_{j=0}^lB_{j,0}( \lambda -E)^{j}
+\sum\limits_{j=0}^{l-1} B_{j,1}( \lambda - E)^{j+{1\over 2}}+
O((\lambda -E)^l),\tag A4.8$$
for $m$ even:
$$R(\l)=\sum\limits_{j=0}^l\sum\limits_{k=0}^\infty
B_{j,k}( \lambda -E)^{j}(\ln(\lambda - E))^k+
+o((\lambda- E)^{l}),\tag A4.9$$
where  $l=0,1,\dots$, $s>C(l,d)$, 
$B_{j,k}  \in B(L_2^s,L_2^{-s} )$,
$B_{j,k}=0$ for 
$k=1$, $j<{d-3\over 2}$ if $d$ is odd and for
$k> {2j\over d-2}$ if  $d$ is even.
Representations {\rm (A4.8), (A4.9)} can be differentiated with
respect to $\l$ any number of times. 
\endproclaim
These results is a standard consequence of the corresponding
properties of the free resolvent (lemmas A4.1,2)
and lemma A4.3,
see [16, 17, 18].
\par Consider the propagator $e^{-itL}$. 
 Lemma A4.4, together with (A4.6), (A4.8), (A4.9)
allows us to represent 
the expression 
$\left<e^{-itL}\hat Pf,g\right>$, $f,g\in C_0^\infty({\Bbb R}^d)$
in the form
$$ \left<e^{-itL}\hat P fg\right>=
\int_E^\infty d\l [e^{-i\l t} \left<{\Cal E}(\l)f,g\right>-
e^{i\l t}\left<{\Cal E}(\l)\s_1f,\s_1g\right>],\tag A4.10 $$
where
$$ {\Cal E}(\l)={1\over 2\pi i}
(R(\l+i0)-R(\l-i0)). $$
It follows from (A4.8), (A4.9) that
as $\lambda\rightarrow E$,
$ {\Cal E}(\l)$ admits the following asymptotic expansion
in $B(L_2^s,L_2^{-s})$ with $s$ sufficiently large.
\newline For $d$ odd: 
$${\Cal E}(\l)=
{\Cal E}_0(\l-E)^{{d-2\over 2}}+
O((\l-E)^{{d\over 2}}),\tag A4.11$$
for $d$ even:
$${\Cal E}(\l)={\Cal E}_0(\l-E)^{{d-2\over 2}}+
\cases O(\ln(\lambda -E)(\lambda-E)^{2})
\quad \roman {if} \,\,\,d=4,\cr
 O((\lambda-E)^{{d\over 2}})
\quad \roman {if}\,\,\, d\geq 6.\cr
\endcases \tag A4.12$$
${\Cal E}_0\in B(L_2^s,L_2^{-s})$.    
These expansions can be differentiated with respect to $\l$
any number of times. 
\par 
Combining (A4.10), (A4.7),  (A4.11), (A4.12)
one gets immediately [18]
$$\|<x>^{-s} e^{-itL}\hat Pf\|_2\leq C<t>^{-d/2}\|<x>^{s}f\|_2,$$
provided $s$ is sufficiently large. To recover proposition 1.1
it is sufficient now to inject  this inequality
in the following representation for $e^{-itL}\hat P$
$$e^{-itL}\hat P =\hat Pe^{-itL_0}
-i\int_0^tdse^{-i(t-s)L_0}\hat PVe^{-isL_0}$$
$$-\int_0^tds\int_s^t
d\rho e^{-i(t-\rho)L_0}Ve^{-i(\rho-s)L}\hat PVe^{-isL_0}.
$$
\vskip10pt

\subhead
Acknowledgement
\endsubhead
It is a pleasure to thank F.Nier
 for numerous helpful discussions.

\subhead
 References
\endsubhead

\vskip8pt
\item{1.} Benjamin, T.B. The stability of solitary waves.
Proc. Roy. Soc. Lond. {\bf  1972}, A{\it 328},  153-183.
\vskip4pt
\item{2.} Berestycki, H.; Lions, P.-L.
Nonlinear scalar field equations, I, II, 
Arch. Rat. Mech. Anal. {\bf 1983}, {\it 82} (4), 313-375.
\vskip4pt
\item{3.} Bourgain, J.; Wang, W. Construction of
blowup solutions for the nonlinear
Schr\"odinger
equation with critical nonlinearity.
Ann. Scuola Norm. Sup. Pisa Cl. Sci. {\bf 1997}, {\it 25} (4), 197-215.
\vskip4pt
\item{4.} Buslaev V.S.; Perelman, G.S. Scattering for the
nonlinear Schr\"odinger equation: states close to  a soliton.
St. Petersburg Math. J. {\bf 1993}, {\it 4} (6),1111-1143.
\vskip4pt
\item{5.} Buslaev,V.S.; Perelman, G.S. On the stability
of solitary waves for nonlinear Schrodinger equation.
Amer. Math. Soc. Transl.(2). {\bf 1995}, {\it 164}, 75-99. 
\vskip4pt
\item{6.} Buslaev, V.S.; Sulem, C. 
 On the asymptotic stability stability
of solitary waves of   nonlinear Schrodinger equations.
Preprint.
\item{7.} Cazenave, T.; P.-L.Lions, P.-L. Orbital stability
of standing waves for some nonlinear Schr\"odinger equations, 
 Commun. Math. Phys. {\bf 1982}, {\it 85} (4), 549-561.
\vskip4pt
\item{8.} Cuccagna, S. Stabilization of solutions to nonlinear
Schr\"odinger equation, Comm. Pure Appl. Math. {\bf 2001}
{\it 54}, 1110-1145.
\vskip4pt
\item{9.} Cuccagna, S. On asymptotic stability of ground states
of NLS. Preprint.
\vskip4pt
\item{10.} Ginibre, J.; Velo G. On a class of nonlinear 
Schr\"odinger equations I, II. J.Func.Anal. {\bf 1979}, {\it 32}, 1-71.
\vskip4pt
\item{11.} Ginibre, J.; Velo G. On a class of nonlinear 
Schr\"odinger equations III. Ann. Inst. H.Poincare -Phys. Theor.
{\bf 1978},
{\it 28} (3), 287-316.
\vskip4pt
\item{12.} Grikurov, V. Preprint, 1995.

\vskip4pt
\item{13.} Grillakis, M.; Shatah, J.; Strauss, W. Stability of 
solitary waves in presence
of symmetry I. J. Func. Anal. {\bf 1987}, {\it 74} (1), 160-197.
\vskip4pt
\item{14.} Grillakis, M.; Shatah J.; Strauss, W. Stability of 
solitary waves in presence
of symmetry II.  J. Func. Anal. {\bf 1990}, {\it 94} (2), 308-384.

\vskip4pt
\item{15.} Hagedorn, G. Asymptotic completeness for
the impact parameter approximation to three particle scattering.
Ann. Inst. Henri Poincar\'e. {\bf 1982}, {\it 36} (1), 19-40. 
\vskip4pt
\item{16.} Jensen, A. Spectral properties of 
Schr\"odinger operators and time decay of the wave functions. 
Results in $L_2({\Bbb R}^m)$, $m\geq 5$.  Duke Math. J.
{\bf 1982},
{\it 47} (1), 57-80.
\vskip4pt
\item{17.} A.Jensen, A. Spectral properties of 
Schr\"odinger operators and time decay of the wave functions. 
Results in $L_2({\Bbb R}^4)$. J. Math. Anal. Appl. {\bf 1984}, 
{\it 101} (2),
397-422.
\item{18.} A.Jensen, A.; Kato, T.
Spectral properties of 
Schr\"odinger operators and time decay of the wave functions. 
  Duke Math. J. {\bf 1979}, {\it 46} (3), 583-611.
\vskip4pt
\item{19.} Martel, Y.; F.Merle, F.; Tsai, T.-P.
Stability and asymptotic stability in the energy space
of the sum of $N$ solitons for subcritical gKdV equations.
Comm. Math. Phys. {\bf 2001}, {\it 231}, 347-373.
\vskip4pt
\item{20.} McLeod, K. Uniqueness of positive radial
solutions of $\triangle u+ f(u)=0$ in ${\Bbb R}^n$.
Trans. Amer. Math. Soc.  {\bf 1993}, {\it 339} (2), 495-505.  
\vskip4pt
\item{21.}
Nier, F.; Soffer, A.
Dispersion and Strichartz estimates for some finite rank
    perturbations of the Laplace operator.
J. of Func. Analysis, to appear.
\vskip4pt
\item{22.} Novikov S.P.(ed.), Theory of solitons: The inverse 
scattering method, Moscow, Nauka, 1980.
\vskip4pt 
\item{23.} Perelman, G. Some results on the scattering of weakly
interacting solitons for nonlinear Schr\"odinger equation.
In: Spectral Theory, Microlocal Analysis, Singular Manifolds,
M.Demuth et al., eds., Math. Top. 14, Berlin, Akademie Verlag, 1997, 
pp. 78-137. 
\vskip4pt 
\item{24.} Pillet, C.-A.; Wayne, C.E.Invariant manifolds
for a class of dispersive, Hamiltonian, partial
differential equations. J. Diff. Eq.  {\bf 1997}, {\it 141} (2), 310-326.
\vskip4pt 
\item{25.}
 Reed, M.; Simon, B.
Methods of modern mathematical physics II: Scattering theory,
New York, Academic Press, 1979.
\vskip4pt 
\item{26.} Shatah, J; Strauss, W. Instability of nonlinear
bounded states. Commun. Math. Phys. {\bf 1987}, {\it 100} (2), 35-108.
\vskip4pt
\item{27.}
 Soffer A.; Weinstein, M.I. Multichannel nonlinear 
 scattering theory for nonintegrable equations I.
Commun. Math. Phys. {\bf 1990}, {\it 133} (1), 119-146.
\vskip4pt
\item{28.} Soffer A.; Weinstein, M.I. Multichannel nonlinear 
 scattering theory for nonintegrable equations II.
J. Diff. Eq. {\bf 1992}, {\it 98} (2), 376-390.
\vskip4pt
\item{29.} Sulem, C.; Sulem, P.-L. 
The nonlinear Schr\"odinger eqation.
Self-focusing and wave collapse,
Appl.Math. Sciences, 139. Springer, New Yowrk, 1999.
\vskip4pt
\item{30.} Tsai, T.-P.; H.-T. Yau, H.-T.
Asymptotic dynamics of nonlinear
Schr\"odinger equations: resonance dominated and
dispersion dominated solutions. Comm. Pure Appl. Math. {\bf 2002},
55, 153-216.
\vskip4pt
\item{31.}  Tsai, T.-P.; Yau,  H.-T. Relaxation of exited states
in nonlinear Schr\"odinger equations. IMRN, to appear.
\vskip4pt
\item{32.} Tsai,  T.-P.; Yau, H.-T.
Stable directions for exited states
of  nonlinear
Schr\"odinger equations. Comm. PDE, to appear.
\vskip4pt
\item{33.} Tsai,  T.-P.; Yau, H.-T. Classification of asymptotic
profiles for nonlinear Schr\"odinger equations
with small initial data. Preprint.
\vskip4pt
\item{34.} Weinstein, M.I. Modulation stability of ground states of 
nonlinear 
Schr\"odinger equations.  SIAM J. Math. Anal. {\bf 1985}, {\it 16} (3), 
472-491.
\vskip4pt
\item{35.} Weinstein, M.I. Lyapunov stability of ground states of 
nonlinear dispersive evolution equations.
 Comm. Pure Appl. Math. {\bf 1986}, {\it 39} (1), 51-68.
\vskip4pt
\item{37.} Weder, R. Center manifold for nonintegrable
nonlinear 
Schr\"odinger equations on the line. Comm. Math. Phys. {\bf 2000},
{\it 215} (2), 343-356.
\vskip4pt
\item{38.}
Yajima, K.
The $W^{k,p}$ continuity of wave operators
for Schr\"odinger operators. J. Math. Soc. Japan {\bf 1995},
{\it 47} (3), 551-581.
\vskip4pt
\item{39.}
Yajima, K.
The $W^{k,p}$ continuity of wave operators
for Schr\"odinger operators. III. Even- dimensional cases
$m\geq 4$. J. Math. Scien. Univ. Tokyo {\bf 1995}, {\it 2} (20, 311-346.

\end